\documentclass[11pt]{article}

\usepackage{bridges}
\usepackage{amsfonts,amssymb,amsthm,eucal,amsmath}
\usepackage{graphicx}
\usepackage{subfig}

\usepackage{wrapfig}
\usepackage{pinlabel, color}
\usepackage{multirow}

\usepackage{microtype}
\usepackage{hyperref}

\usepackage[para]{manyfoot}
\DeclareNewFootnote[para]{B}

\newcommand{\CC}{\mathbb{C}}

\newcommand{\RR}{\mathbb{R}}
\newcommand{\TT}{\mathbb{T}}
\newcommand{\ZZ}{\mathbb{Z}}

\newcommand{\calI}{\mathcal{I}}

\newcommand{\CP}{\mathbb{CP}}

\newcommand{\SC}{\mathfrak{sc}}

\newcommand{\RS}{\widehat{\mathbb{C}}}

\newcommand{\Imagein}{\calI_\text{in}}
\newcommand{\Imageout}{\calI_\text{out}}

\newcommand{\reffig}[1]{Figure~\ref{Fig:#1}}

\title{Squares that Look Round: Transforming Spherical Images\thanks{This work is in the public domain.}}
\renewcommand\footnotemark{}

\author{
 \begin{tabular}{cc}
  Saul Schleimer & Henry Segerman \\
  Mathematics Institute & Department of Mathematics \\
  University of Warwick & Oklahoma State University
 \end{tabular}
}

\date{}

\begin{document}
\maketitle

\begin{abstract}
  We propose M\"obius transformations as the natural rotation and
  scaling tools for editing spherical images.  As an application we
  produce spherical Droste images.  We obtain other self-similar
  visual effects using rational functions, elliptic functions, and
  Schwarz-Christoffel maps.
\end{abstract}

%%% ArXiv classes - cs.GR, math.HO
%%% AMS classes - 00A66 (visualization), 33C75 (elliptic integrals), 30F10 (uniformization)
%%% ACM classes - I.3.3

\section*{Introduction}
\label{Sec:Intro}

Interest in spherical imagery has grown in recent years, driven by
increased availability of both viewing devices and cameras.  The
YouTube application on smartphones now plays spherical video, using
the phone's accelerometer.  On the camera side, numerous
consumer-focused spherical cameras are available, as well as high-end
professional offerings.

\begin{wrapfigure}[21]{r}{0.42\textwidth}
\centering
\vspace{-15pt}
\includegraphics[width=0.39\textwidth]{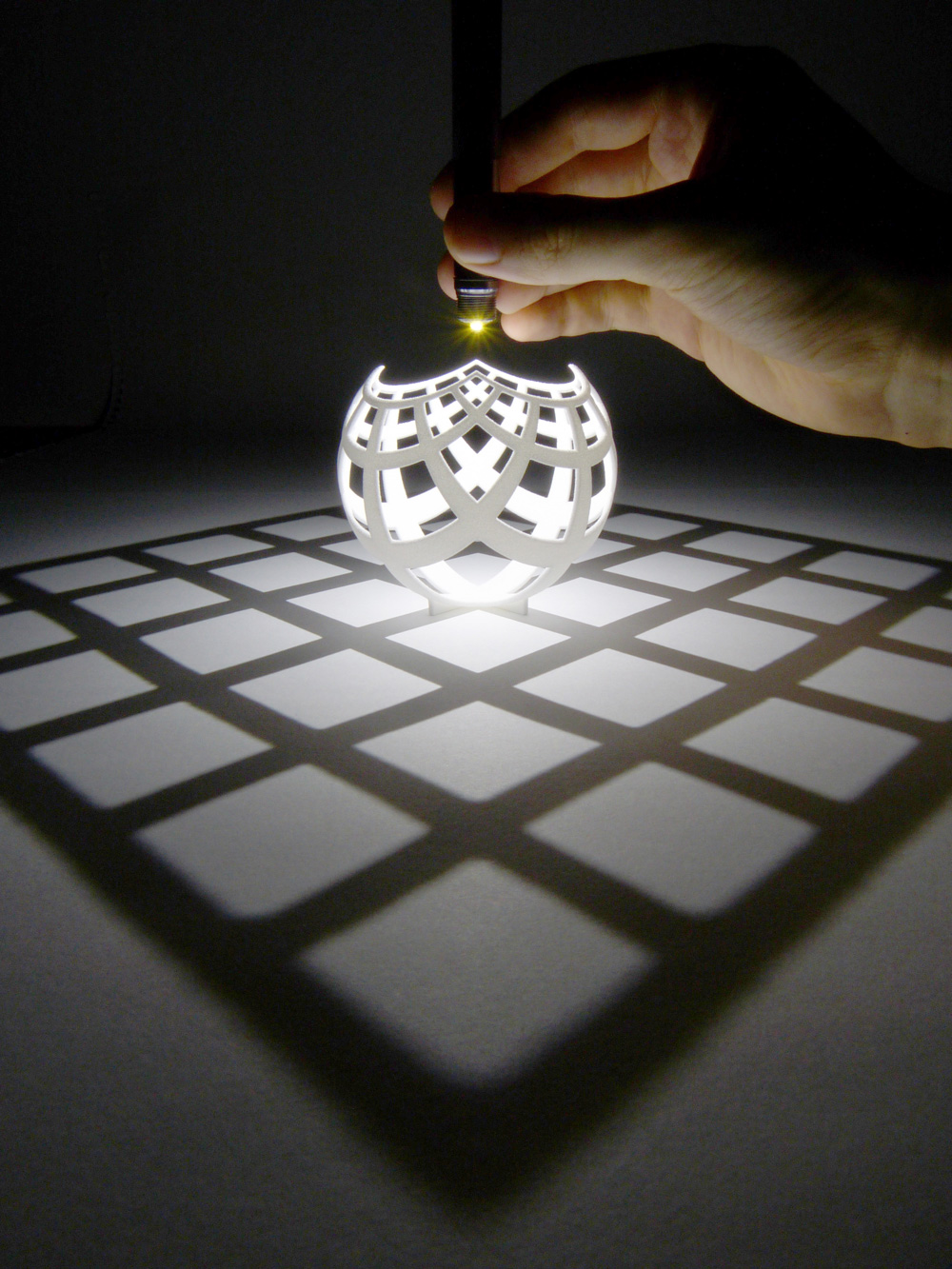}
\caption{Stereographic projection from the sphere to the plane.}
\label{Fig:Stereo_proj}
\end{wrapfigure}

Almost universally, spherical images and video are stored and
transmitted via \emph{equirectangular projection}: points on the
sphere are given by their latitude and longitude.  Thus the whole
image is stored as a rectangular image with a two-to-one aspect ratio,
corresponding to angles $(0,2\pi) \times (-\pi/2, \pi/2)$.  This data
format fits conveniently into the existing infrastructure for ordinary
images.  However, there is a problem: most tools for editing ordinary
rectangular images, when applied to the equirectangular projection,
give poor results.  For example, standard rectangular editing tools cannot
rotate a spherical image about a non-vertical axis.

Future editing tools for spherical images will no doubt include the
ability to rotate images around any axis, giving analogues of both
translation and rotation of flat images.  However, we can also ask
what scaling (in video, zooming) might mean for spherical images.  In
this paper, we first recall how M\"obius transformations naturally
rotate and scale the sphere.  We then use these to produce spherical
Droste images.  We also obtain other interesting visual effects using
rational functions, elliptic functions, and Schwarz-Christoffel maps.

\vspace{2\lineskip}
\noindent
\textbf{Acknowledgements:} The second author was inspired
by a paper of S\'ebastien P\'erez-Duarte and David
Swart~\cite{bridges2013:217}.  See also \cite{new_methods}.  He began
work whilst visiting \textit{eleVR} (a research group consisting of
Emily Eifler, Vi Hart and Andrea Hawksley) and wrote a guest blog
post\footnote{\url{http://elevr.com/spherical-video-editing-effects-with-mobius-transformations/}}
explaining the implementation of M\"obius transformations.  The Python
code used to generate many of these images is available at
\textit{GitHub}\footnote{\url{https://github.com/henryseg/spherical_image_editing}}.
All spherical photographs and videos were taken using a Ricoh Theta S.

\section*{M\"obius transformations}
\label{Sec:Mob}

M\"obius transformations act on the \emph{Riemann sphere}, $\RS = \CC
\cup \{\infty\}$. This is the result of adding a single point,
$\infty$, to the complex plane $\CC$. We map from the unit sphere
$S^2$ in $\RR^3$ to the Riemann sphere using stereographic
projection~\cite[page~57]{MumfordEtAl02}:
\[  
\rho(u, v, w) = \frac{u + iv}{1 - w}
\]
We set $\rho(0,0,1) = \infty$. Every other point of the unit sphere
maps to a point of $\CC$.  \reffig{Stereo_proj} shows a 3D printed
visualisation of stereographic projection.
%%% (or, strictly speaking, stereographic projection composed with
%%% scaling the plane by a factor of two -- if the plane of the table
%%% went through the center of the sphere, then this would show
%%% stereographic projection as defined above).
A \emph{M\"obius transformation} $M = (a, b; c, d)$ is the map from
the Riemann sphere to itself given by
\[
M(z) = \frac{az + b}{cz + d},
    \quad \mbox{where $a,b,c,d \in \CC$ and $ad - bc \neq 0$.}
\]
There are various special cases involving the point at infinity.  If
$cz + d = 0$ then $M(z) = \infty$.  If $c \neq 0$ then $M(\infty) =
a/c$.  If $c = 0$ then $M(\infty) = \infty$.  There is a cleaner
definition, avoiding these special cases, which uses the
one-dimensional complex projective space $\CP^1$ which we use in our
implementation.  Here, to simplify the exposition, we use $\RS$.

%%% There is no canonical reference for CP^1, so we can't give one...

%%% CP^1: the set of ordered pairs $(z,w)$ of complex numbers, where
%%% we say that $(z,w)=(\lambda z, \lambda w)$ for every
%%% $\lambda\in\CC-\{0\}$.  A M\"obius transformation is the
%%% result of applying a $2\times2$ matrix of complex numbers to the
%%% vector $(z,w)$, with the condition that the determinant of the
%%% matrix is non-zero.

\begin{figure}[htbp]
\centering
\subfloat[]
{
\includegraphics[width=0.22\textwidth]{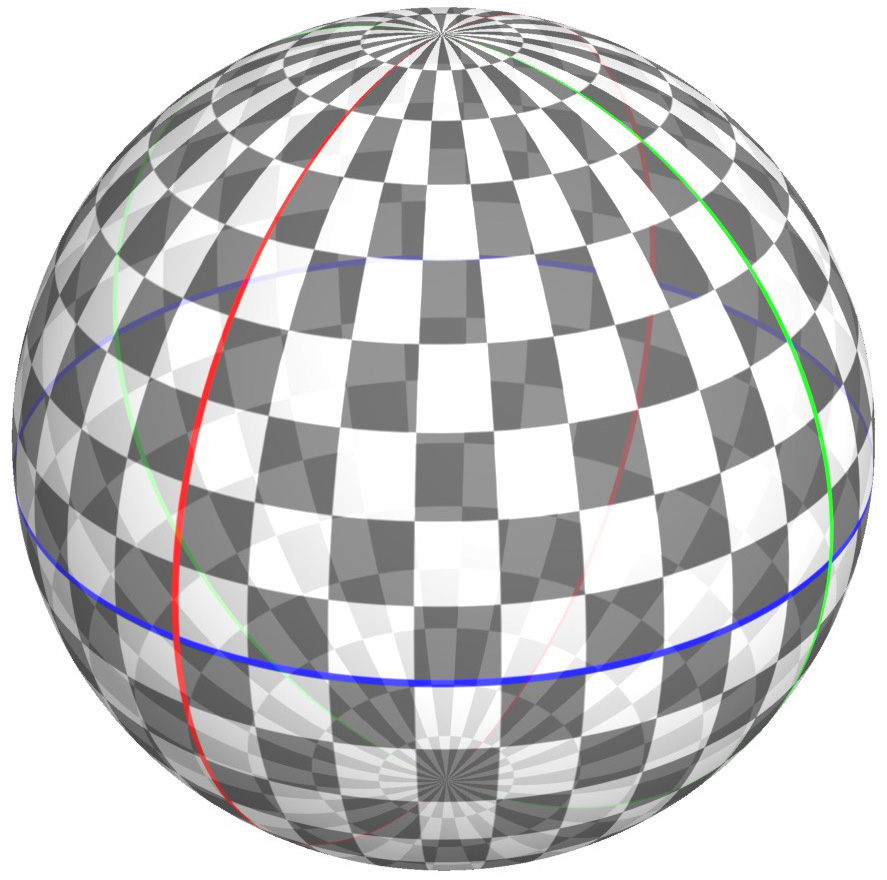}
\label{Fig:test_pattern_sphere}
} 
\quad
\subfloat[]
{
\includegraphics[width=0.22\textwidth]{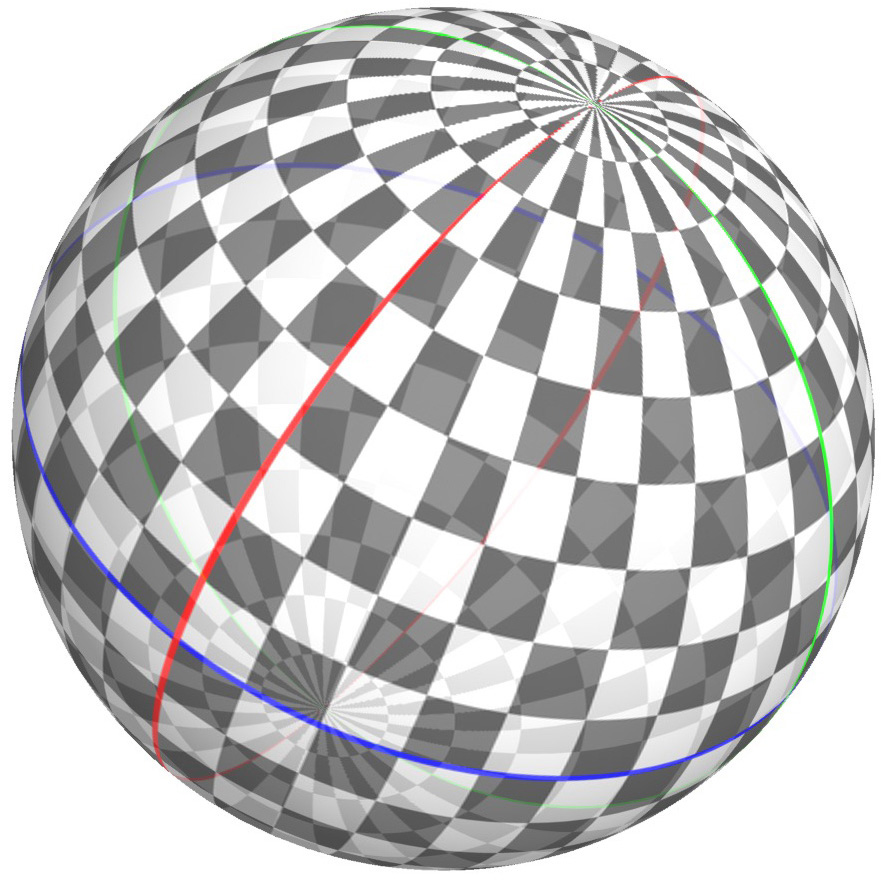}
\label{Fig:rotated_test_pattern_sphere}
} 
\quad
\subfloat[]
{
\includegraphics[width=0.22\textwidth]{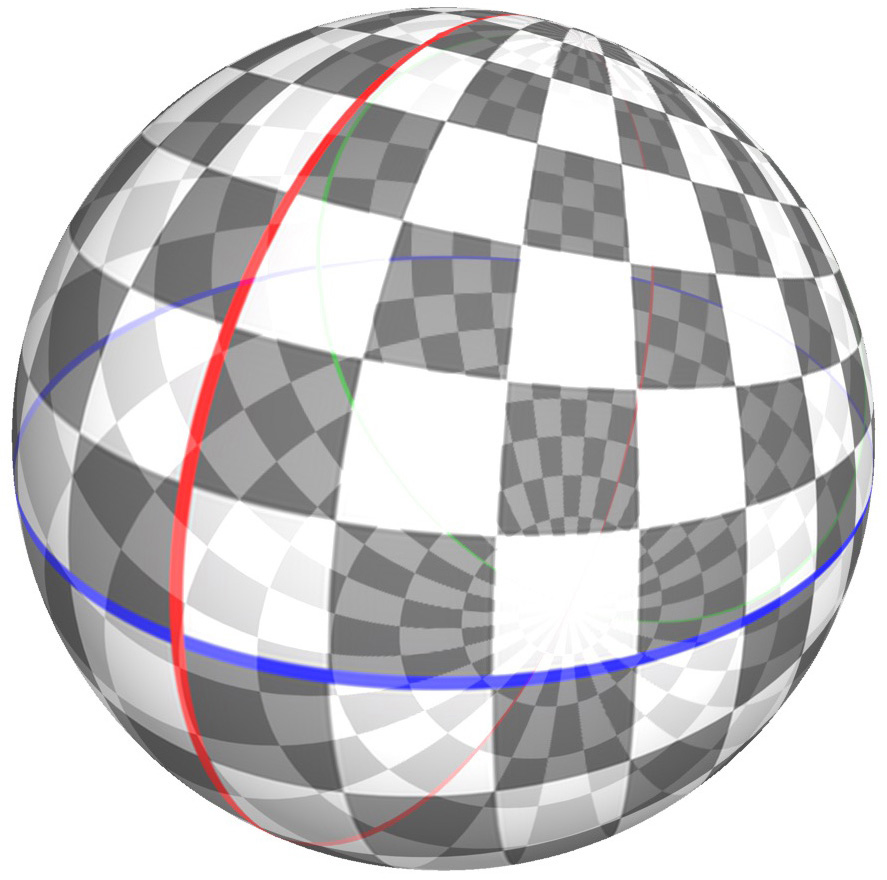}
\label{Fig:scaled_test_pattern_sphere}
} 
\quad
\subfloat[]
{
\includegraphics[width=0.22\textwidth]{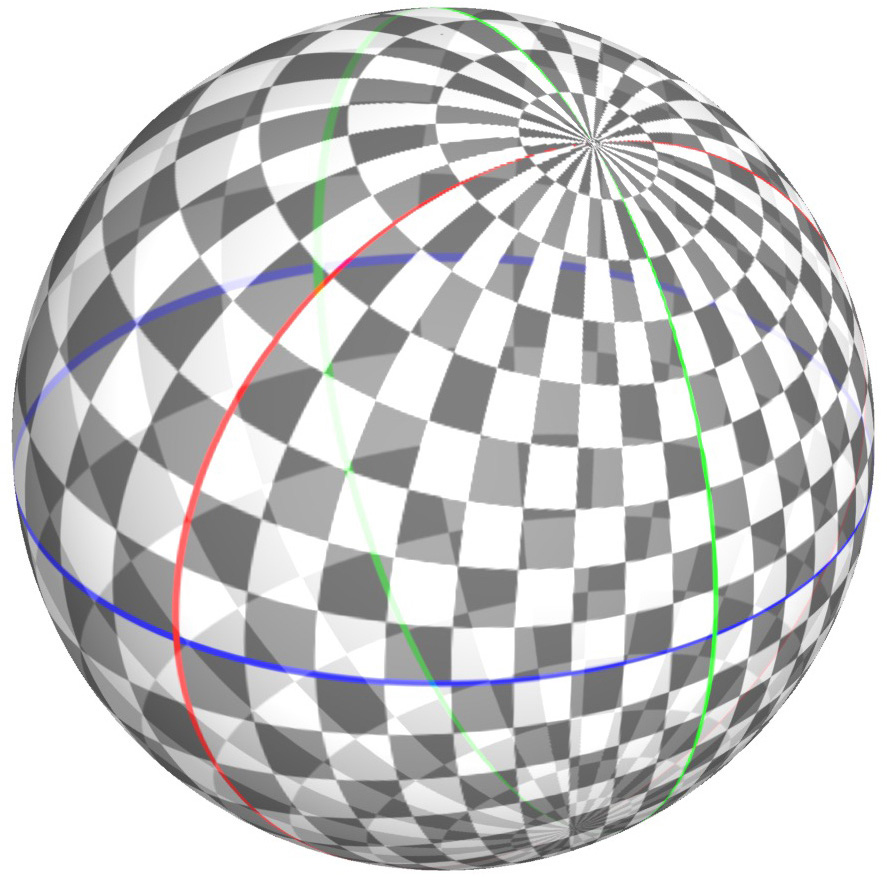}
\label{Fig:parab_test_pattern_sphere}
} 

\subfloat[]
{
\includegraphics[width=0.22\textwidth]{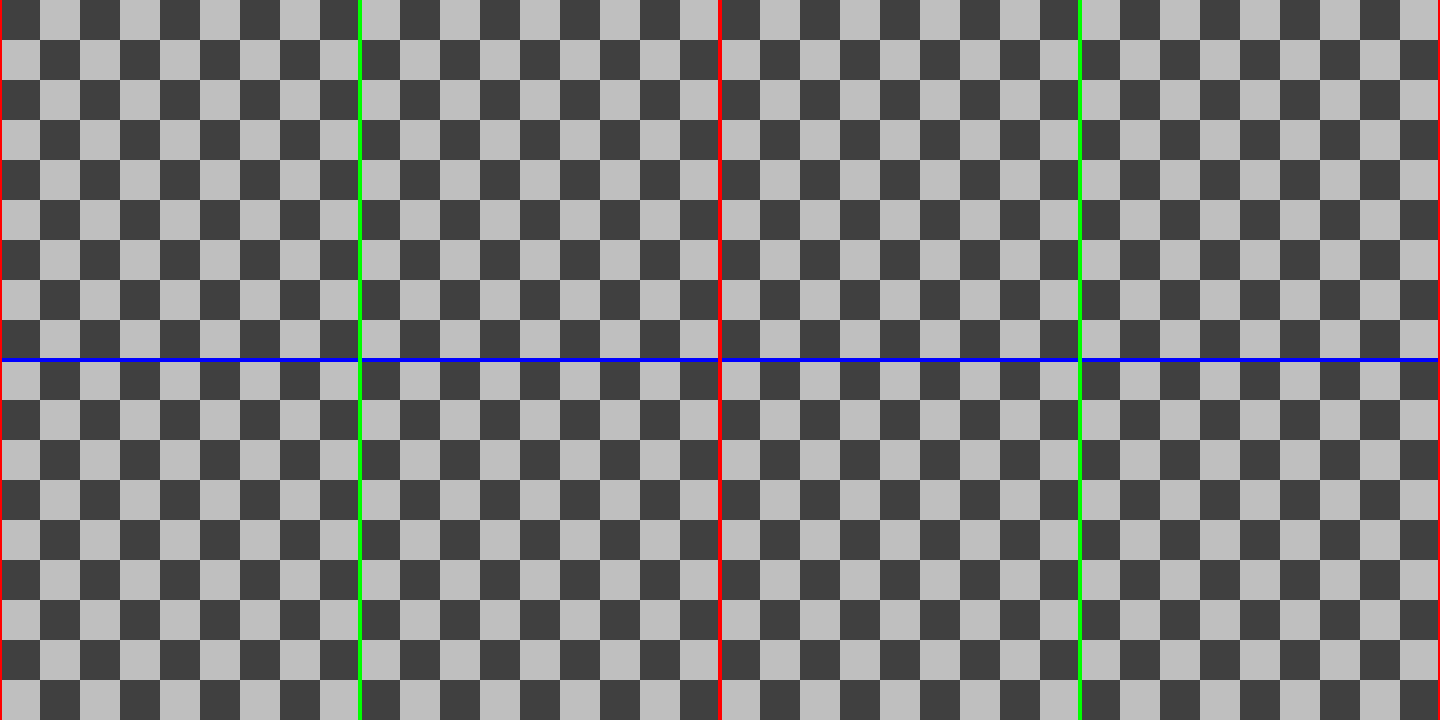}
\label{Fig:test_pattern}
} 
\quad
\subfloat[]
{
\includegraphics[width=0.22\textwidth]{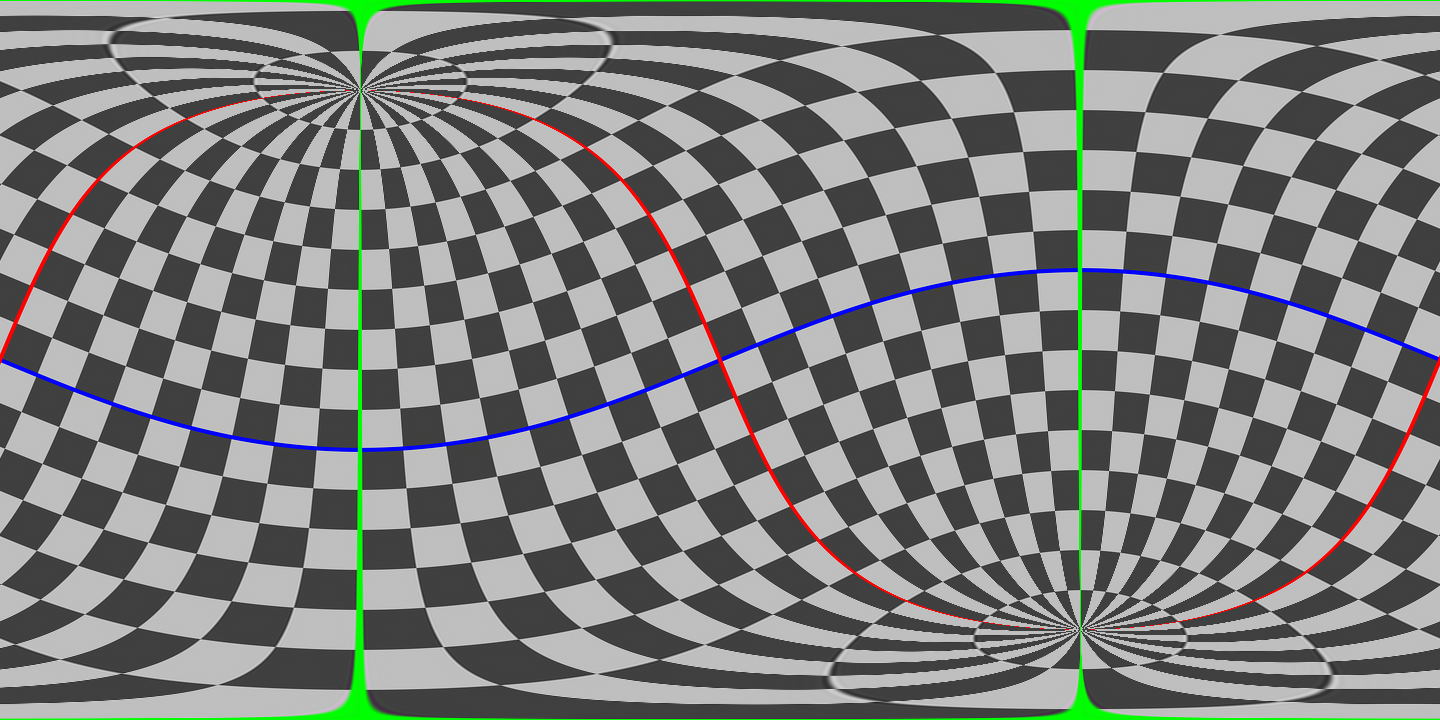}
\label{Fig:rotated_test_pattern}
} 
\quad
\subfloat[]
{
\includegraphics[width=0.22\textwidth]{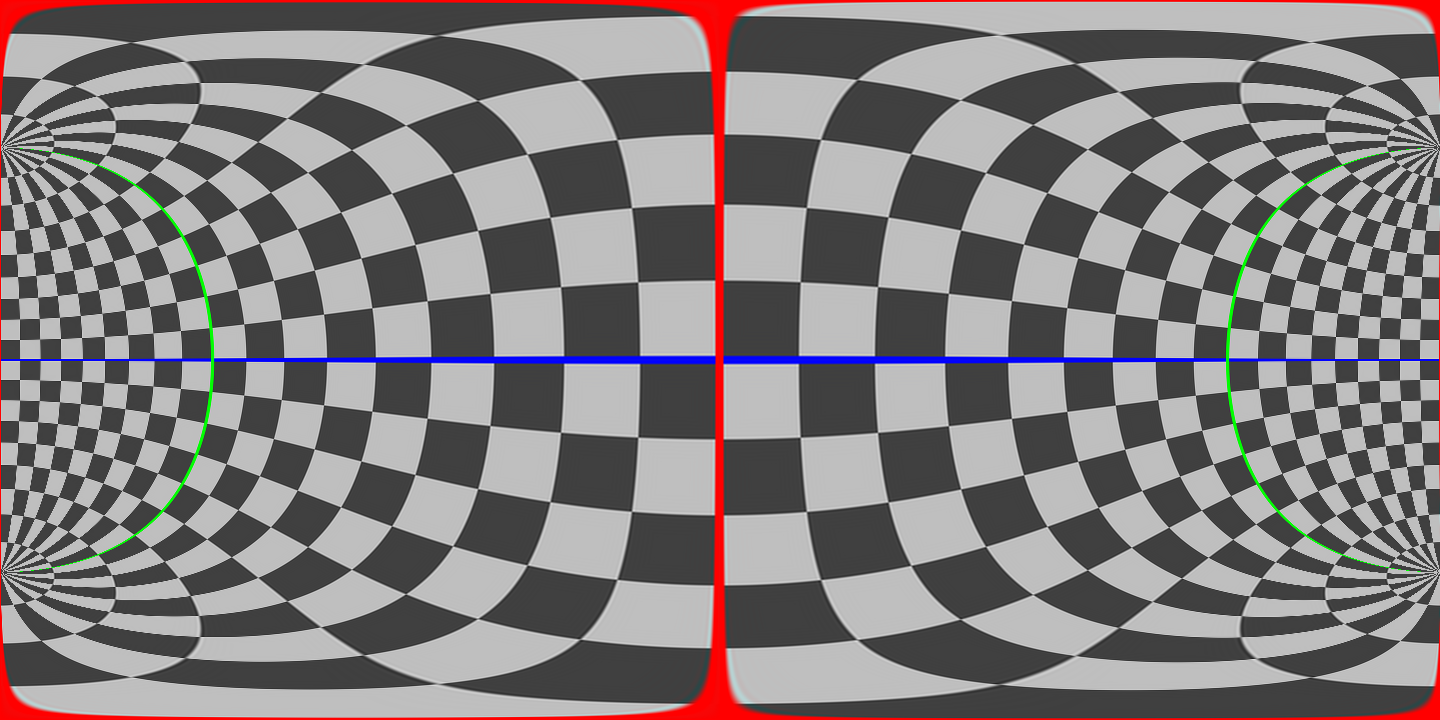}
\label{Fig:scaled_test_pattern}
} 
\quad
\subfloat[]
{
\includegraphics[width=0.22\textwidth]{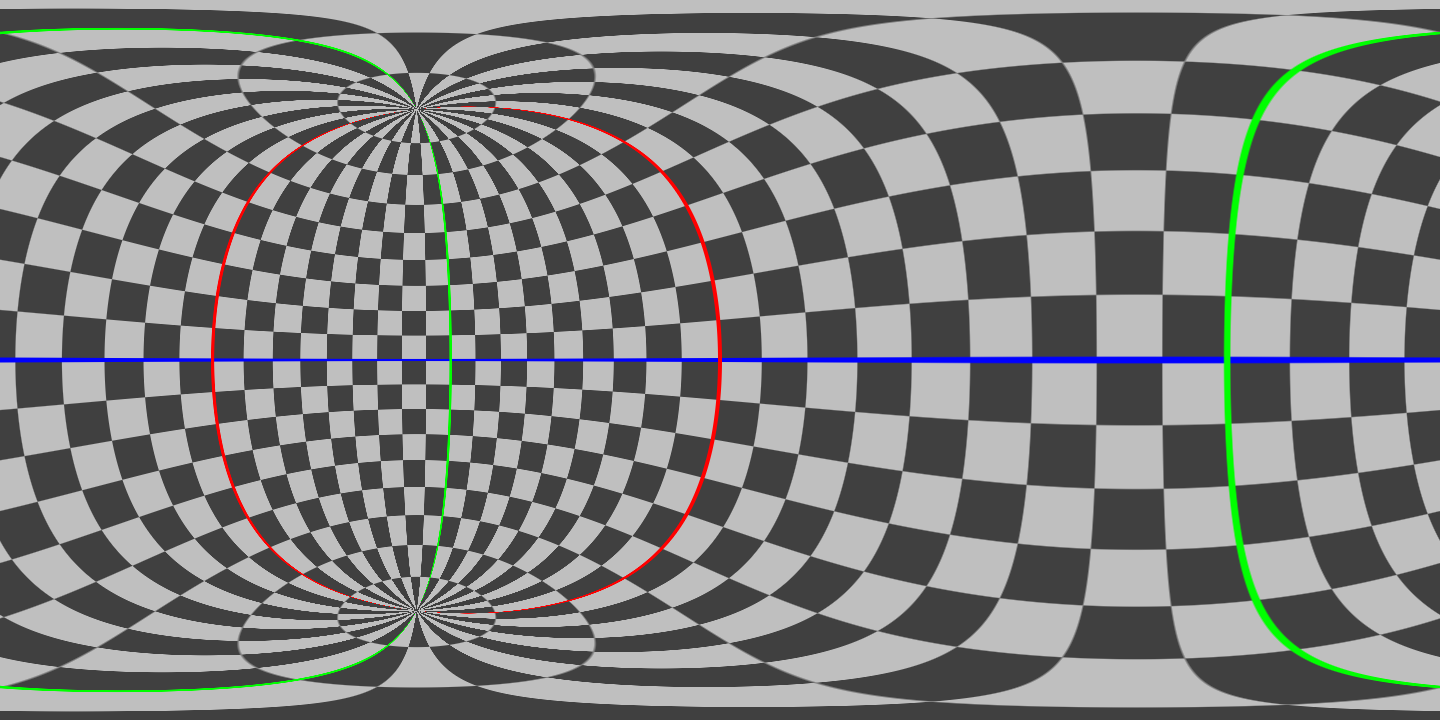}
\label{Fig:parab_test_pattern}
}
\caption{A test pattern (a and e), and the results of rotating by
  $\pi/8$ (b and f), scaling by a factor of two (c and g) and applying
  the parabolic translation $M(z)=z+1/2$ (d and h). Above: the
  textures on the sphere. Below: their equirectangular
  projections. Note that we generally view a spherical image from
  inside the sphere. From this perspective the equirectangular
  projections have the same orientation as the textures on the
  spheres.}
\label{Fig:mob_tsfms}
\end{figure}

We can rotate the complex plane about $0$ by multiplying by a unit
complex number, say $e^{i\theta}$.  We can scale the plane, again
centered on $0$, by multiplying by a real number, say $\lambda \in
\RR$.  Finally, we can translate the plane by adding a complex number,
say $w$.  These give $M = (e^{i\theta}, 0; 0, 1)$ (\emph{elliptic}),
$M = (\lambda, 0; 0, 1)$ (\emph{hyperbolic}), and $M = (1, w; 0, 1)$
(\emph{parabolic}).  Every M\"obius transformation is equivalent, via
conjugation, to one of these.

\begin{figure}[htbp]
\centering

\subfloat[The input image.]
{
\includegraphics[width=0.48\textwidth]{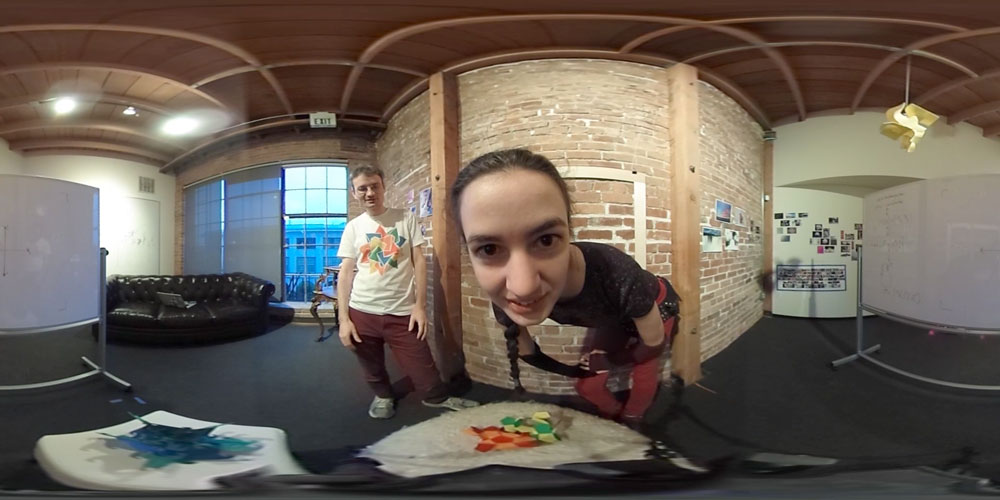}
\label{Fig:Vi_eye}
} 
\subfloat[The result of rotating by an angle of $\pi/12$.]
{
\includegraphics[width=0.48\textwidth]{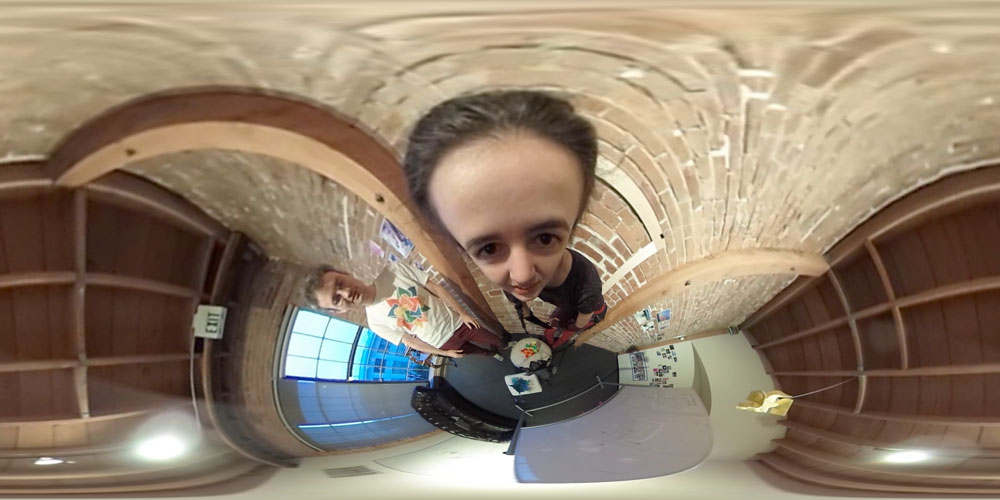}
\label{Fig:Vi_eye_rotate}
} 

\caption{Rotating a spherical photograph of Vi Hart and Henry
  Segerman, about Vi's eyes.}
\label{Fig:Vi_eye_rotate_main}
\end{figure}

\reffig{mob_tsfms} (top row) shows an initial test pattern, and
the results of applying a rotation by $\theta = \pi/8$, of scaling by
a factor of $\lambda = 2$, and of adding $w = 1/2$.  The parabolic
case is included for completeness; it is not clear how this might be
used in image editing.
%%% Hmm - what two finger gestures might we use to implement these?
%%% Rotate and pinching are already common.  So for parabolics I
%%% suggest ``place the thumb and index finger and then, keeping the
%%% thumb fixed, move the index finger in the desired direction.'' :)
%%%
%%% Arnaud C. says that the only reasonable gestures are three-finger
%%% gestures, using the fact that PSL(2, C) is three-transitive.
Here we have placed zero, the origin of the complex plane, at the
``front pole'' of the sphere: the front intersection of the blue
equator and red longitude.  Note how the hyperbolic transformation
scales distances up by a factor of two at zero but scales distances
\emph{down} by a factor of two at the antipodal point, $\infty$.  In
fact, M\"obius transformations allow us to rotate or scale fixing any
two points of the sphere.  As an example, see
\reffig{Vi_eye_rotate_main}; we show a frame of raw footage and the
transformed frame from a
video\footnote{\url{https://www.youtube.com/watch?v=oVwmF_vrZh0}}
exploring many of the effects in this paper.

Every M\"obius transformation, other than rotations about antipodal points, distorts spherical distance. However, as illustrated in 
\reffig{mob_tsfms}, right angles always remain right angles.  In fact, M\"obius
transformations are \emph{conformal}: they preserve all angles.  Thus
images are not sheared or non-uniformly stretched; features remain
essentially recognisable.  All of the transformations in this paper
mapping the sphere to itself are conformal, apart from at a discrete
set of points.
%%% Omit: {conformal, invertable} <-> {M\"obius}
%%% Pf: Postcompose with a M\"obius transformation so that
%%% the composition fixes 0, 1, \infty.  Then the restriction to the
%%% plane is conformal, fixes 0, 1, and is invertable. ... //
Note that the equirectangular projection is not conformal; both
distances and angles are distorted.

% Conventions for labelling the equirectangular projection:

% (0)---(0)---(0)---(0)---(0)
%  |     |     |     |     |
%  |     |     |     |     |
% (1)---(i)--(-1)--(-i)---(1)
%  |     |     |     |     |
%  |     |     |     |     |
% inf---inf---inf---inf---inf

% This is a right-handed coordinate system.  Yay!

\section*{Pulling back, pushing forward, and branch points}
\label{Sec:PullPush}

If we want to apply a transformation $T$ to a pixel-based input image
$\Imagein$, we need to find the inverse transformation $S = T^{-1}$.
%%% Pixel graphics are a ``function from'', vectors graphics are a
%%% ``function to''.
This is because the algorithm to generate the output image $\Imageout$
runs in reverse: for each desired pixel $p$ of $\Imageout$, we take
its position $z_p$, compute $S(z_p)$, and assign $p$ the same color as
the pixel with position $S(z_p)$ in $\Imagein$.  (In fact we take a
weighted average of colors of input pixels nearest to $S(z_p)$.)
%%% That is, ``anti-aliasing''.
Note that, in order to have an algorithm, the transformation $S =
T^{-1}$ must be single-valued, but $T$ need not be.
%%% If the deriviative of $S$ is large at $z$ then we may want to take
%%% an average of colors of input pixels to avoid a jagged appearance
%%% in the output.
We call this procedure \emph{pulling back} via $S$ or, equivalently,
\emph{pushing forward} via $T$.
%%% Finding S = T^{-1} is easy for Mobius transformations, but not so
%%% much for more interesting functions...

\begin{figure}[htbp]
\centering
\subfloat[The input image.]
{
\includegraphics[width=0.48\textwidth]{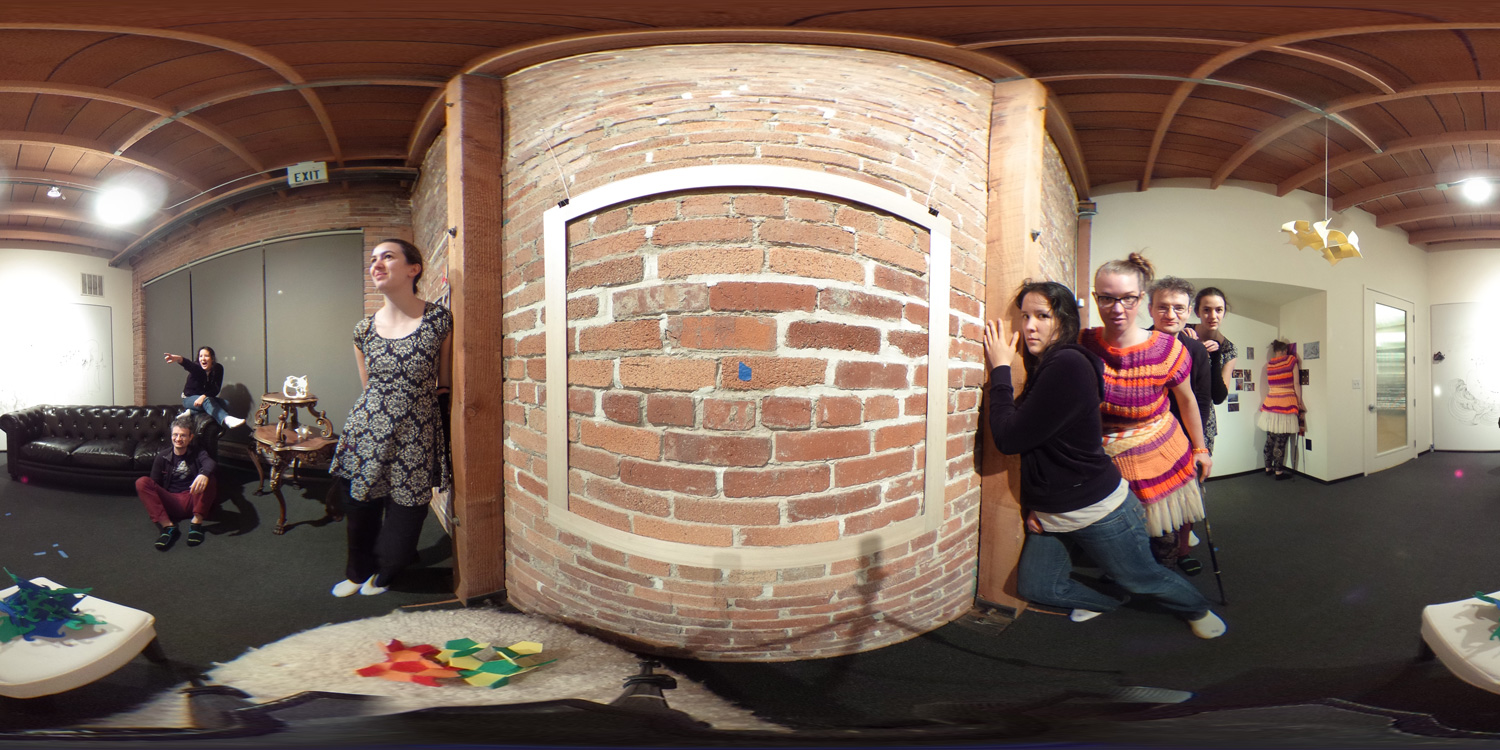}
\label{Fig:elevr_image}
}
% So, we have:
% 0 = ceiling
% inf = tripod
% 1 = back of room
% i = Vi
% -1 = blue tape
% -i = Emily
\subfloat[Pulling back by $S(z) = z^2$.]
{
\includegraphics[width=0.48\textwidth]{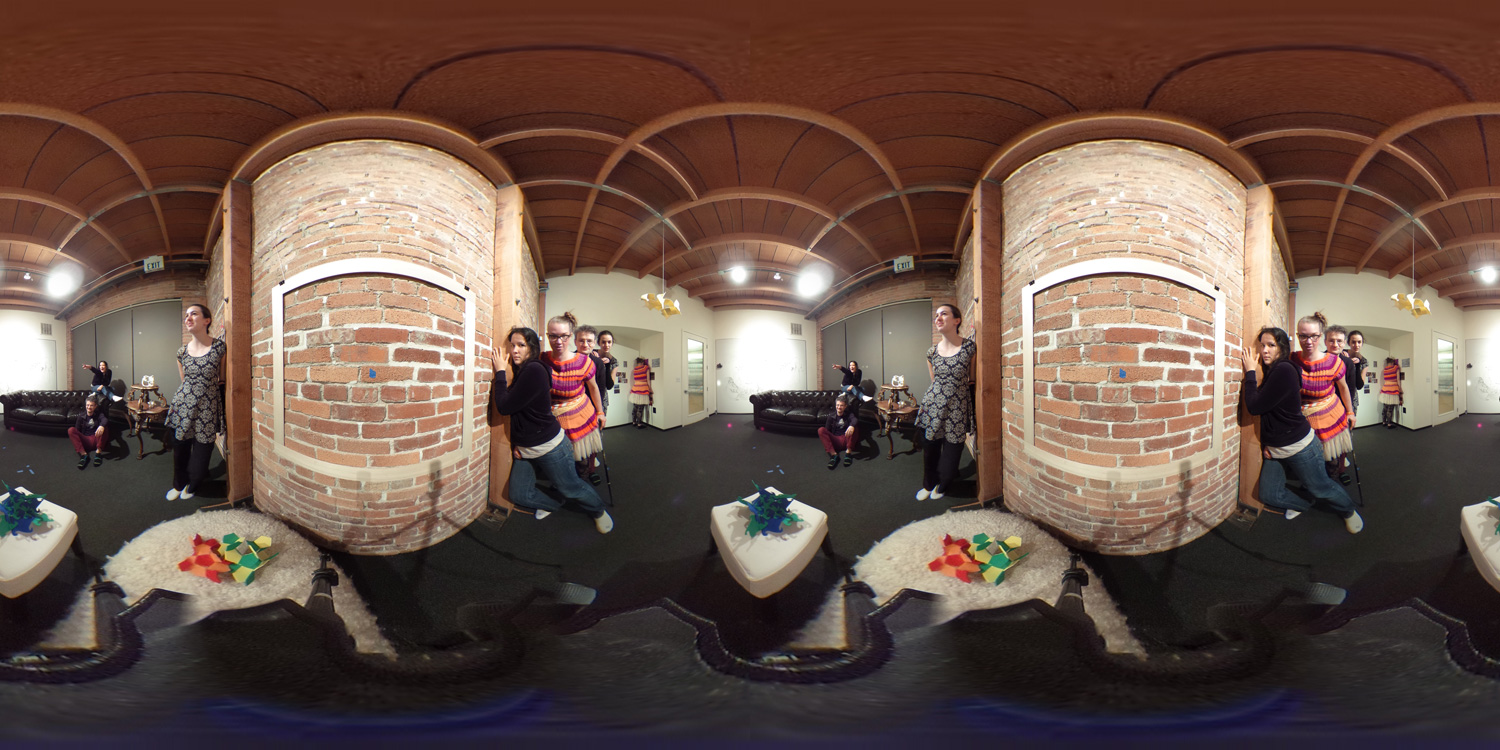}
\label{Fig:z_squared}
} 

\subfloat[Rotated to show an order two branch point at the center.]
{
\includegraphics[width=0.48\textwidth]{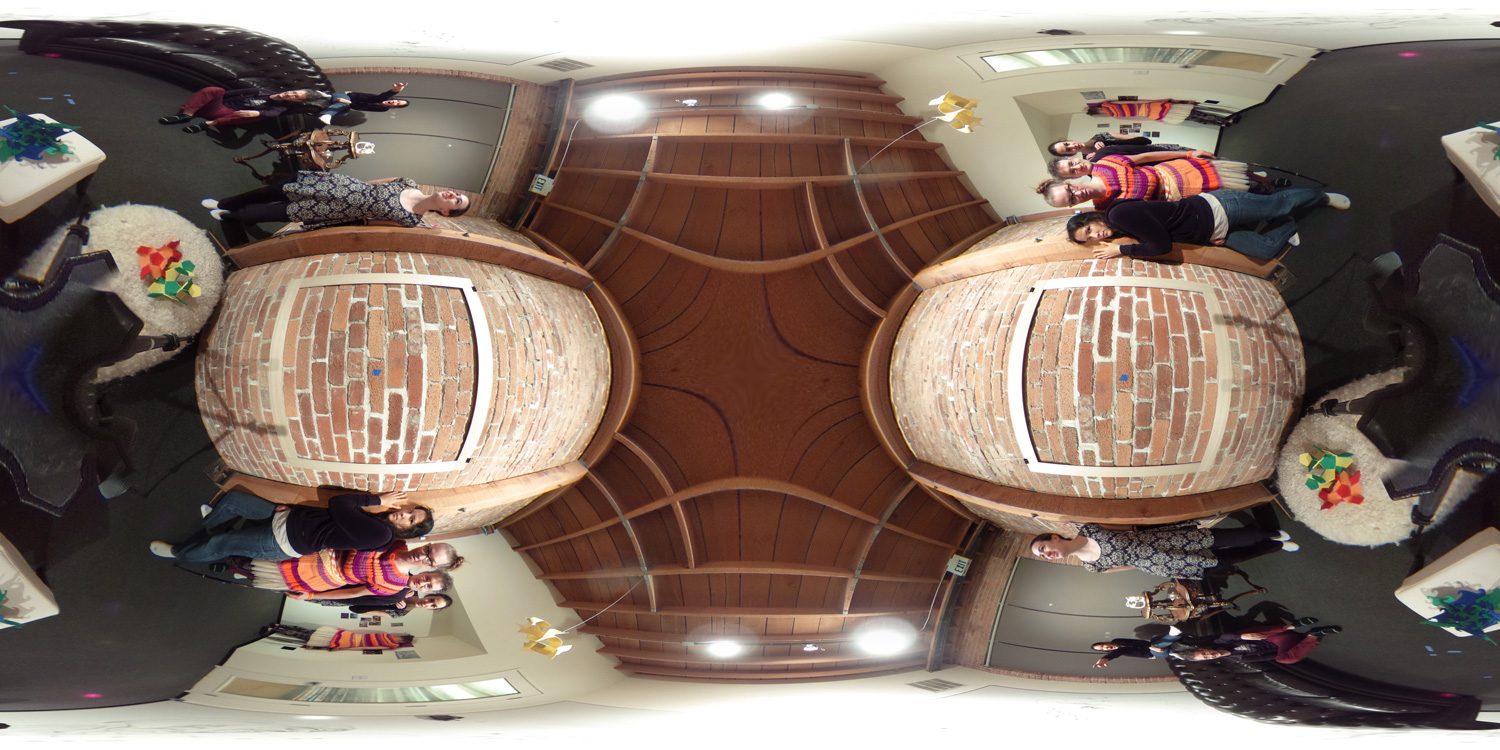}
\label{Fig:order_2_branch_point}
} 
\subfloat[Pulling back via $S(z) = -e^{-4\left(\frac{1+z}{1-z}\right)}$.]
{
\includegraphics[width=0.48\textwidth]{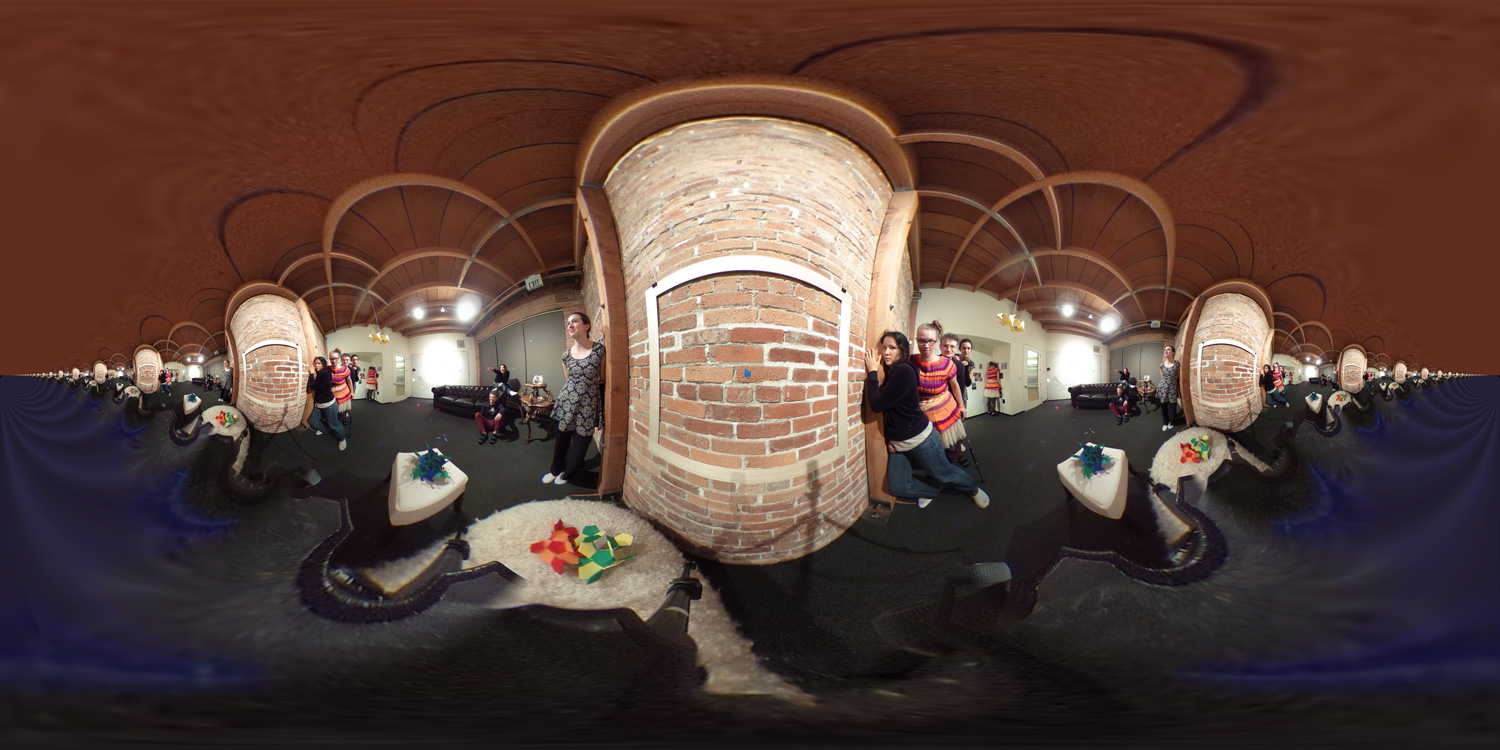}
\label{Fig:exp_variant}
} 

\caption{Transformations applied to a spherical photograph featuring
  Emily Eifler, Vi Hart, Andrea Hawksley, and Henry Segerman, all
  appearing twice. In these images the origin $0 \in \CC$ corresponds to
  the top of the equirectangular projection while infinity corresponds
  to the bottom, other than for \reffig{order_2_branch_point}. }
\label{other_transformations}
\end{figure}

Now consider \reffig{elevr_image}.  If we take $T(z) = \pm \sqrt{z}$
and $S(z) = z^2$ then we obtain \reffig{z_squared}. Here we see a new
feature, \emph{branch points}, around which nearby imagery is
repeated.  To see the branch point more clearly, we rotate
\reffig{z_squared} to get \reffig{order_2_branch_point}.  The number
of repetitions is the \emph{order} of a branch point. In $\RS$ the
branch points are of order two and lie at zero and infinity. In
Figures~\ref{Fig:z_squared} and~\ref{Fig:order_2_branch_point} they
are on the floor and the ceiling.  These branch points are
unavoidable: any conformal transformation of $\RS$ is either a
M\"obius transformation or has branch
points~\cite[Section~4.3.2]{Ahlfors66}.
%%% Exercise 2 or Exercise 4.  

\reffig{exp_variant} shows the result of pulling back via a variant of
the complex exponential map, specifically $S(z) =
-e^{-\lambda\left(\frac{1 + z}{1 - z}\right)}$.  Here $\lambda$ is a
scaling parameter and the M\"obius transformation $M(z) = \frac{1 +
  z}{1 - z}$ is a rotation by $\pi/2$ about $\pm i$; this ensures that
the image repeats horizontally rather than vertically.
%%% The minus sign ensures that up is up. :)
In this case, the forward transformation $T(z)$ is a variant of the
complex logarithm, so is infinitely valued.
% To be precise: 
% w = -exp(-4(1+z)/(1 - z))
% -1/4 log(-w) = (z + 1)/(-z + 1)
%%%%    need inverse = (z - 1)/(z + 1)
% z = \frac{log(-w) + 4}{log(-w) - 4}
% z = \frac{log(w) + \pi i + 4}{log(w) + \pi i - 4}
Thus the output contains infinitely many copies of the input image.
The branch points are again on the floor and the ceiling, but are of
infinite order.  

The same techniques can be used to combine different spherical images
into a single spherical image.  This provides a spherical analogue of
the familiar ``split screen'' trope in rectangular video: compositing
multiple video clips into a single screen.  We, however, can stitch
the different images seamlessly, if they match along suitable arcs
between the branch points. We created a spherical video along these
lines, in which the second author appears to be in a two-fold branched
cover of his
apartment\footnote{\url{http://www.youtube.com/watch?v=UUW_ZU3_TQM}}. The
footage is stitched together with a video of the empty apartment, so
that only one copy of the author appears in the combined video.

\section*{The Droste effect}
\label{Sec:Droste}

\begin{figure}[htbp]
\centering

\subfloat[A Droste annulus.]
{
\includegraphics[width=0.33\textwidth]{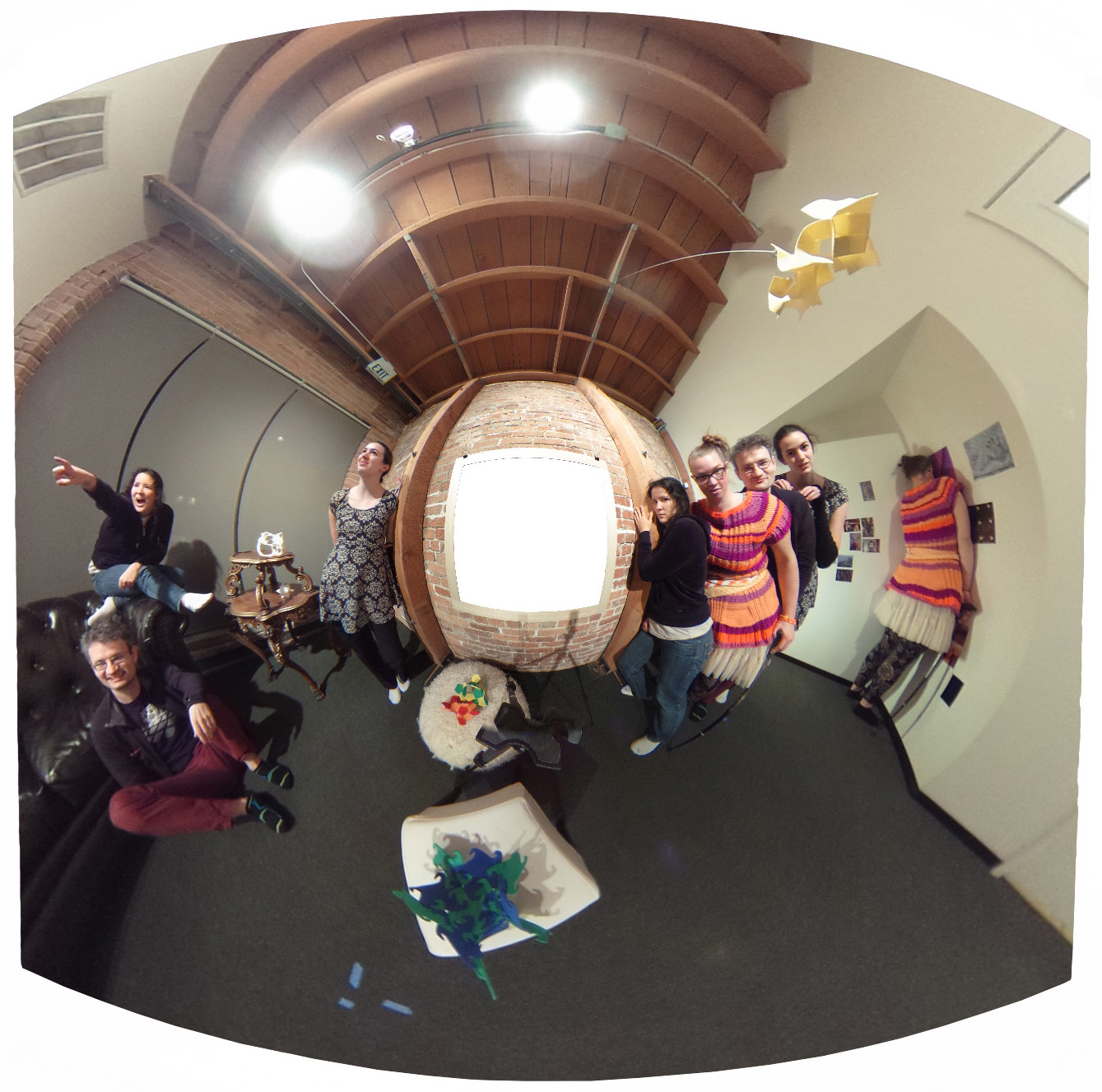}
\label{Fig:spherical_droste_annulus}
} 
\subfloat[A straight Droste image.]
{
\includegraphics[width=0.62\textwidth]{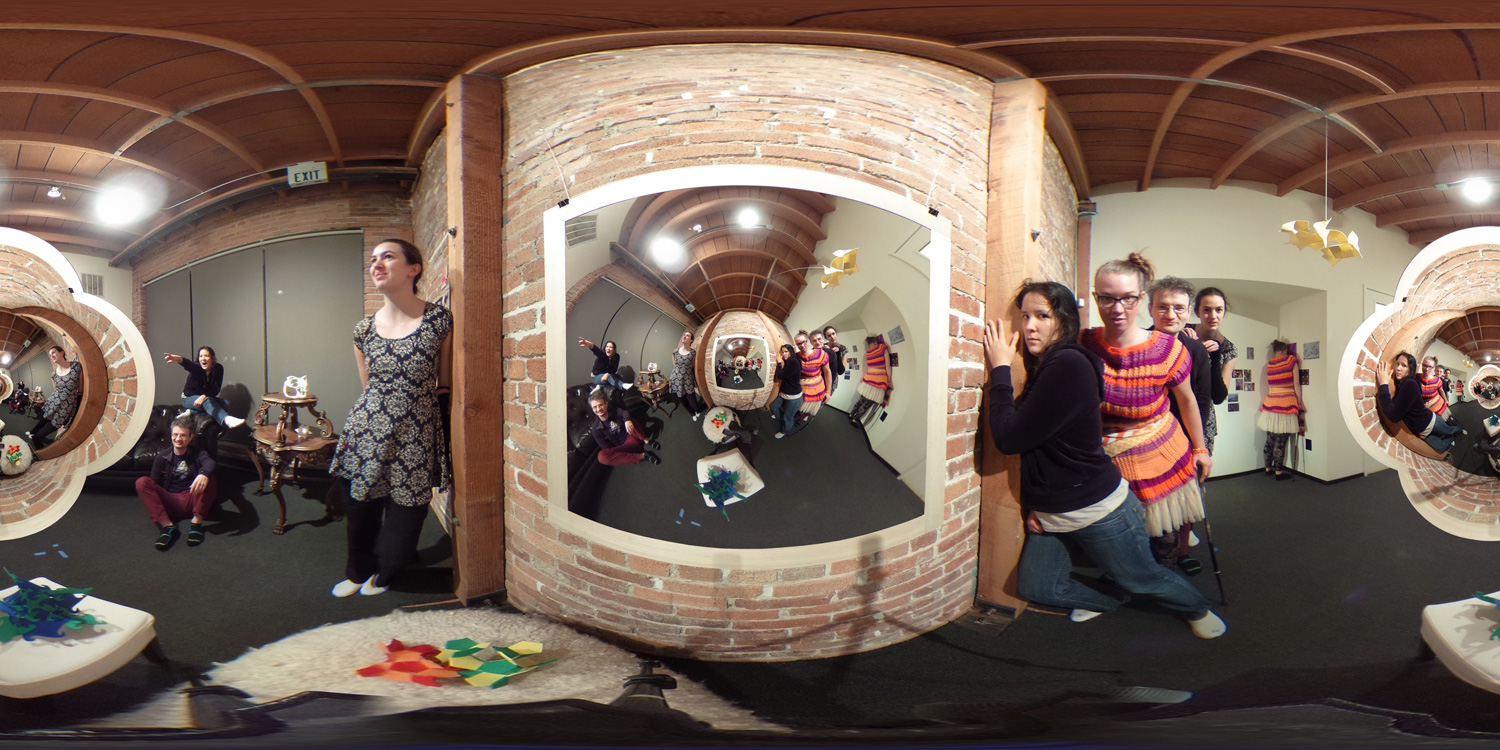}
\label{Fig:spherical_droste_straight}
} 

\subfloat[Log of the Droste annulus.]
{
\includegraphics[height = 4.0in]{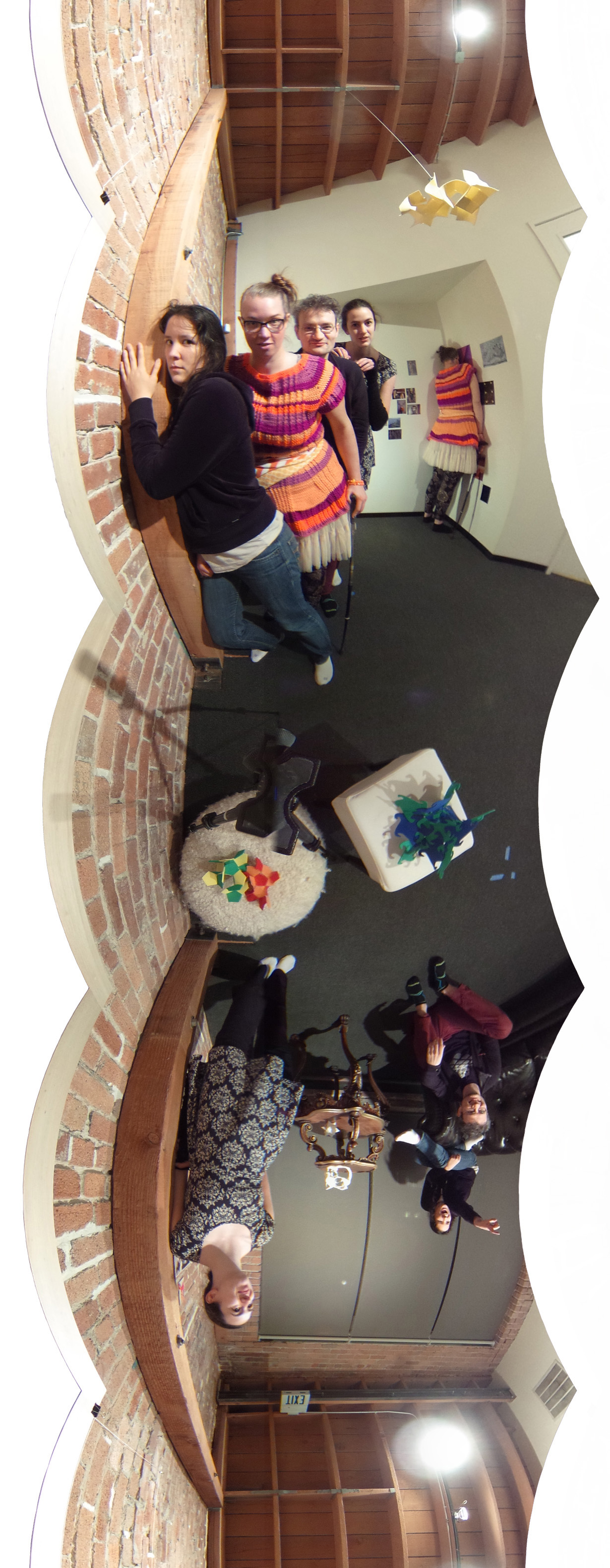}
\label{Fig:spherical_droste_annulus_log}
} 
\qquad
\subfloat[A different fundamental rectangle.]
{
\includegraphics[height = 4.0in]{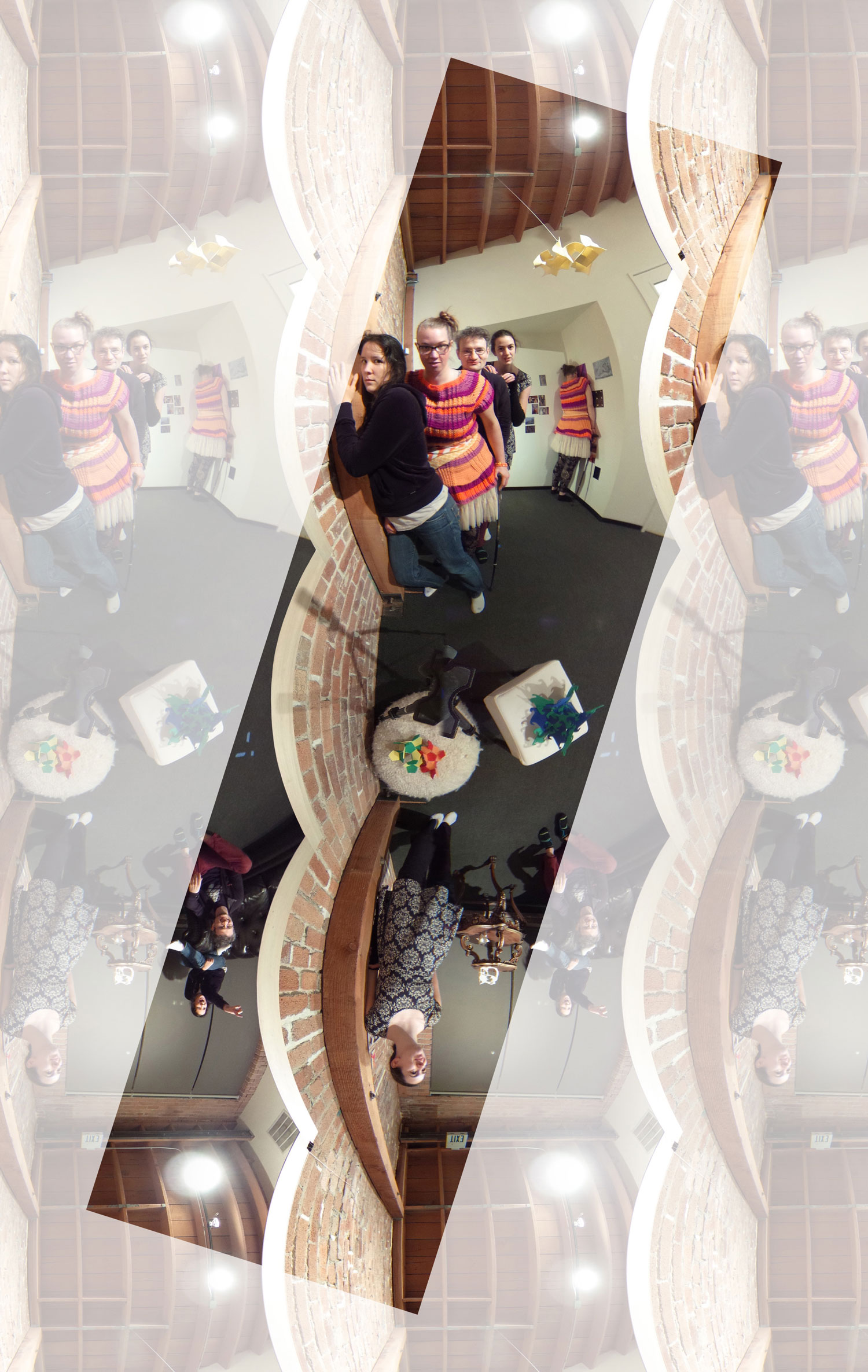}
\label{Fig:spherical_droste_log_overlay_twist}
} 

\subfloat[A twisted Droste image.]
{
\includegraphics[width=0.48\textwidth]{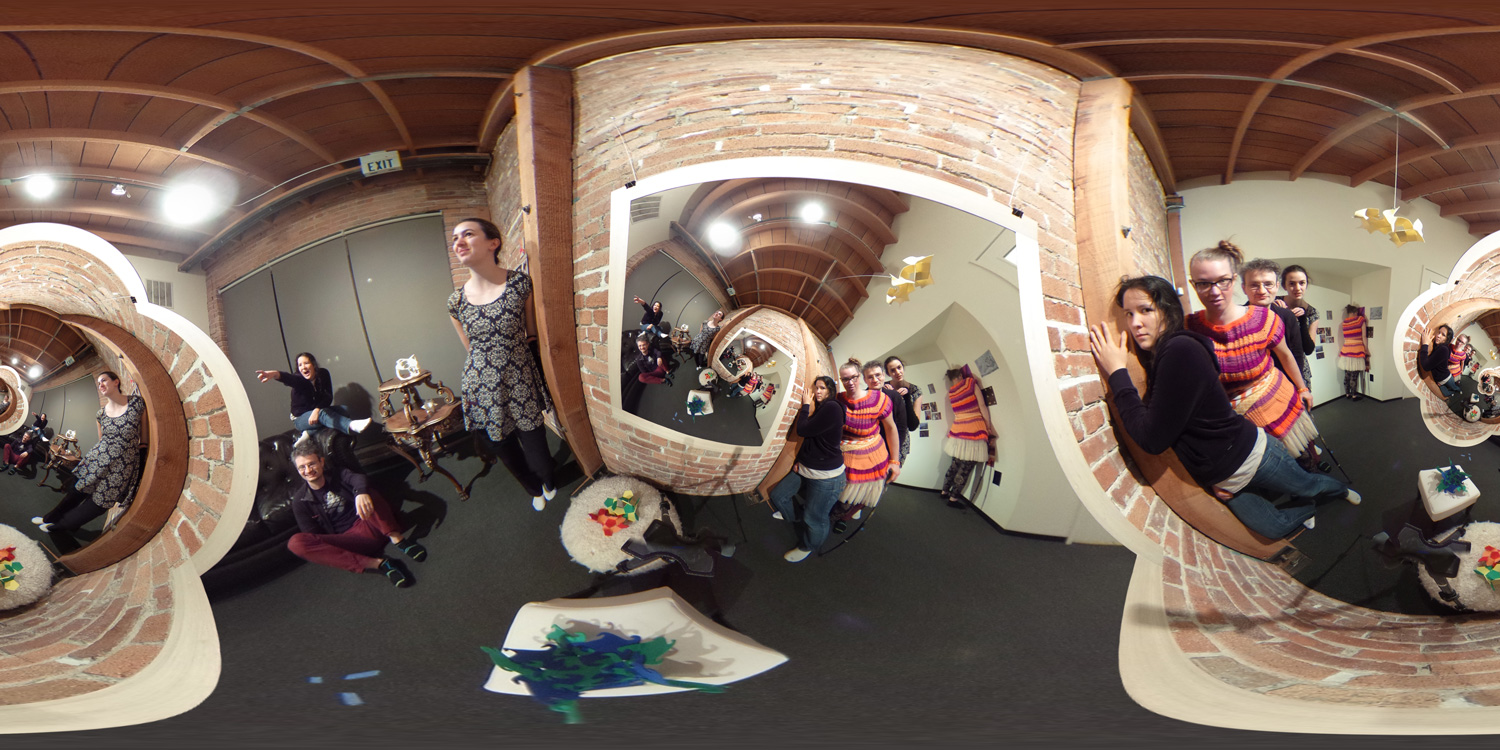}
\label{Fig:spherical_droste_twisted}
}  
\subfloat[Here the image inside the frame differs from the one outside.]
{
\includegraphics[width=0.48\textwidth]{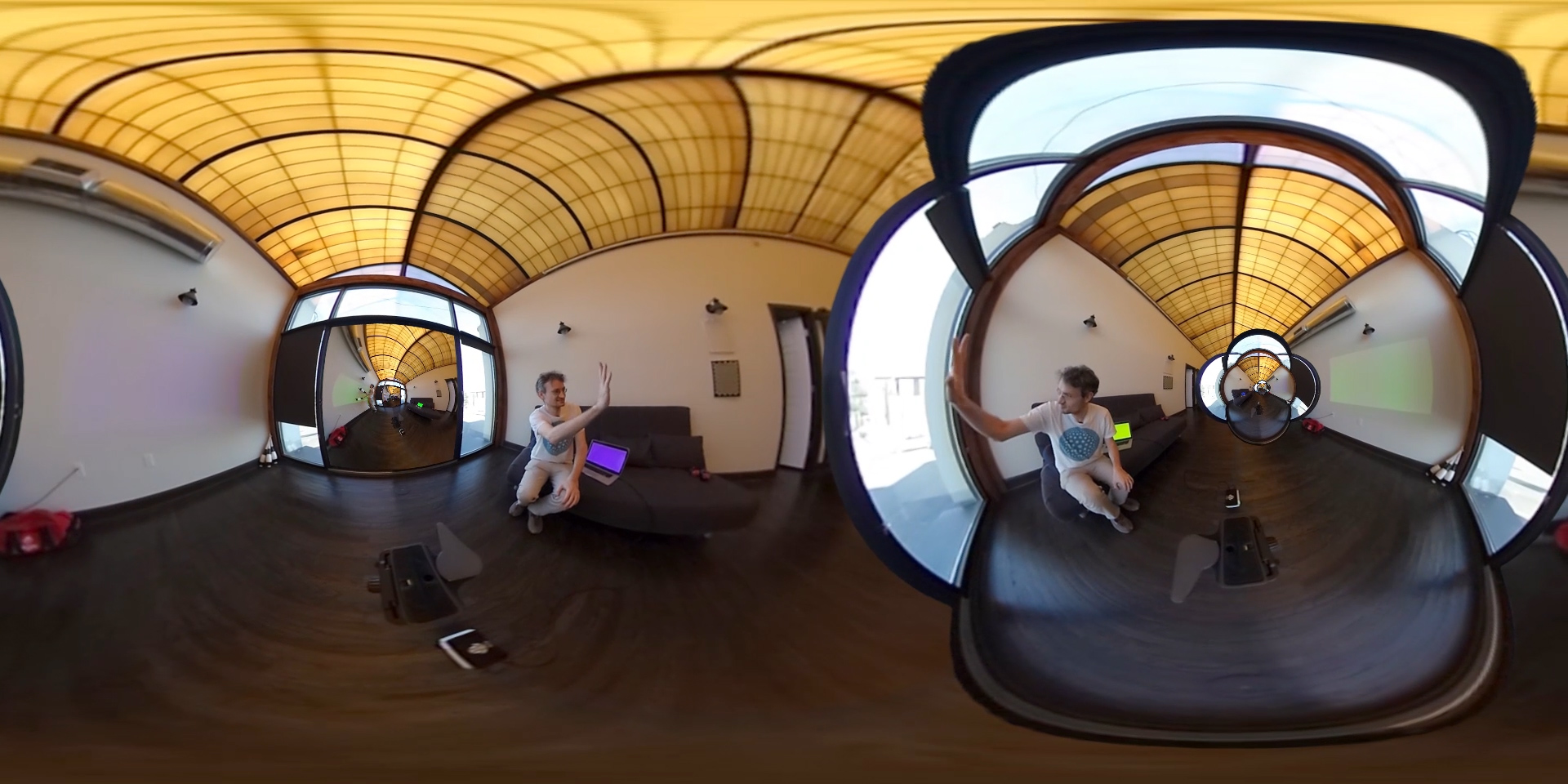}
\label{Fig:spherical_droste_different_images}
} 

\caption{Droste effect images. These images answer the question of
  what an observer inside of a Droste effect image sees when they look
  away from the frame: there is a flower-shaped portal floating in the
  middle of the room.}
\label{Fig:spherical_droste}
\end{figure}

A common artistic and mathematical motif is that of
``self-similarity''; this is often called the \emph{Droste effect} in
commercial and computer graphics.
%%% Droste, land o'lakes, happy cow, ... this is very old - see
%%% wikipedia.  
A ``straight Droste effect'', as found on the packaging of the
eponymous Dutch cocoa, is obtained when the entire picture is
included, under a shrinking transformation, inside of itself.  The
``twisted Droste effect'' was first introduced by M.C.~Escher in his
\emph{Print Gallery} lithograph.  The mathematics behind Escher's
image was explained by Bart de Smit and Hendrik
Lenstra~\cite{escherdroste}.

It is possible to obtain both the straight and twisted Droste effect
in spherical images using M\"obius transformations, the complex
exponential map, and the complex logarithm (see
also~\cite[page~223]{bridges2013:217}).  We simplify the discussion
here by suppressing all mention of equirectangular and stereographic
projections.  We begin with a spherical image, say
\reffig{elevr_image}.
%%% really in equirect coords.  Push to sphere and then to plane.
We remove everything inside a small disk (here the inside of the
frame on the wall) and everything outside a larger disk, to obtain a \emph{Droste
  annulus}; see \reffig{spherical_droste_annulus}.  We arrange matters
so that there is a scaling transformation $M(z) = \lambda z$ that
takes the outer boundary of the annulus to the inner boundary.  Thus
we may tile the sphere (minus two points) by copies of the annulus,
obtaining a straight Droste image; see
\reffig{spherical_droste_straight}.

Applying a logarithm unwraps the Droste annulus to give an infinite
vertical strip in $\CC$ with width $\log \lambda$; see
\reffig{spherical_droste_annulus_log}.  Another way to obtain the
straight Droste effect is to tile the plane by horizontal translations
of the strip and apply the exponential map.
% because exp turns translation by log \lambda into multiplication by
% \lambda.  This will be relevant later.
Instead, we may follow De~Smit and
Lenstra~\cite[Figure~10]{escherdroste}, and obtain a twisted Droste
effect.  We scale and rotate so that the rectangle shown in
\reffig{spherical_droste_log_overlay_twist} is vertical and has height
$2\pi$.  Applying the exponential map yields
\reffig{spherical_droste_twisted}.

\addtocounter{footnoteB}{4}  % get footnote and footnoteB on the same number

\reffig{spherical_droste_different_images} shows a still image from a
straight Droste
video\footnoteB{\url{https://www.youtube.com/watch?v=qvh-EAipIUk}} in
which different Droste annuli are offset in time as well as in
space.  The video is continually scaled, giving the impression of
movement through the window.  The time offset between neighboring
annuli matches the apparent flight time of the camera; thus the video
loops.

All constructions of Droste effect images seem to involve ``cut-and-paste''
techniques;
%%% Ok, is there a proof of this?
here we had the choice of frame and the choice of scaling.  In
contrast, the pullback techniques of the previous section can be
applied to any spherical image whatsoever.  The output is seamless;
the only blemishes are the branch points.

\begin{wrapfigure}[28]{r}{0.35\textwidth}
\vspace{-10pt}
\centering

\subfloat[Peirce's quincuncial projection.]
{
\includegraphics[width=0.35\textwidth]{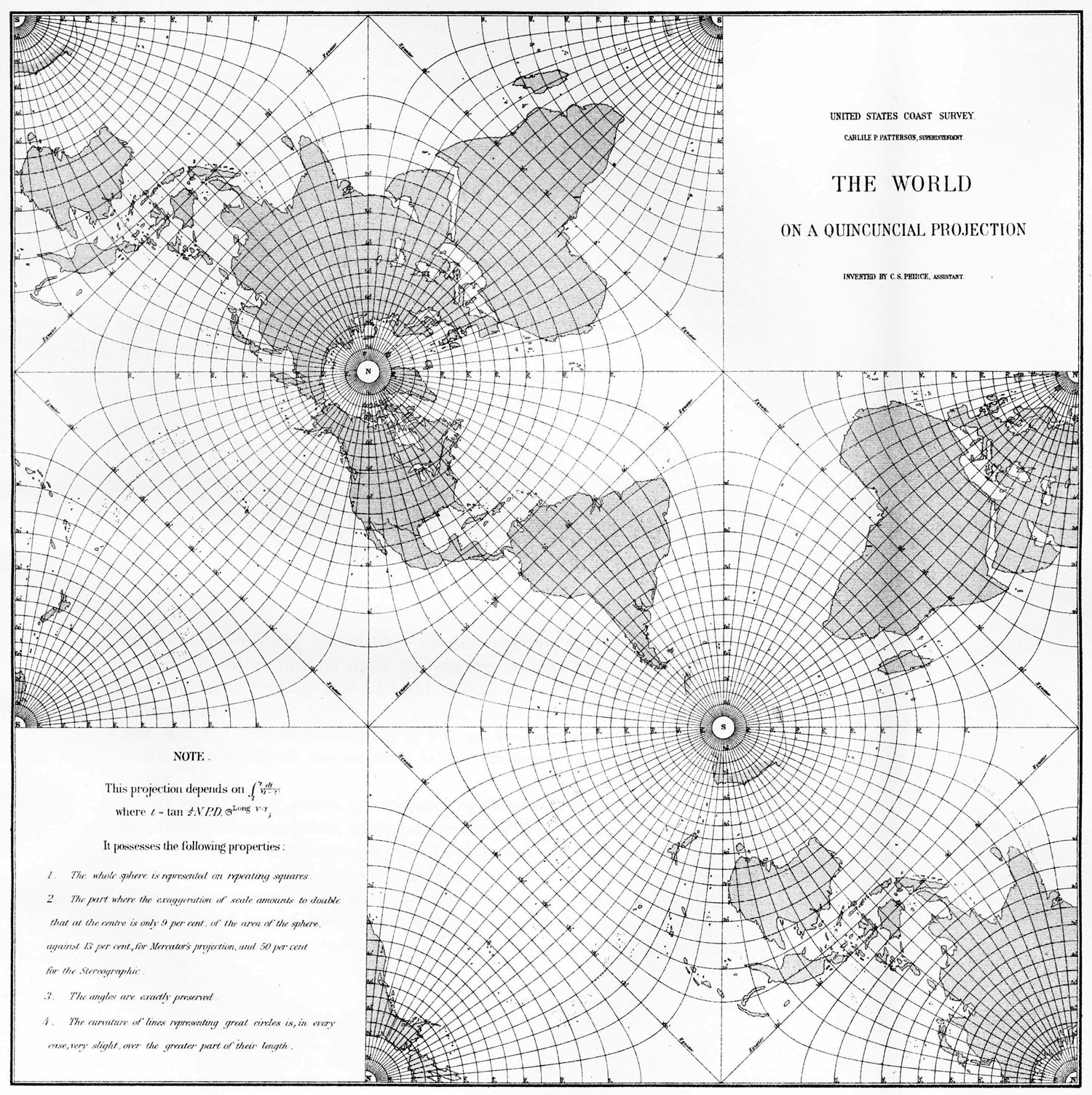}
\label{Fig:Peirce_figure}
} 

\subfloat[On a torus, as in \reffig{sphere_torus_skewer}.]
{
\includegraphics[width=0.35\textwidth]{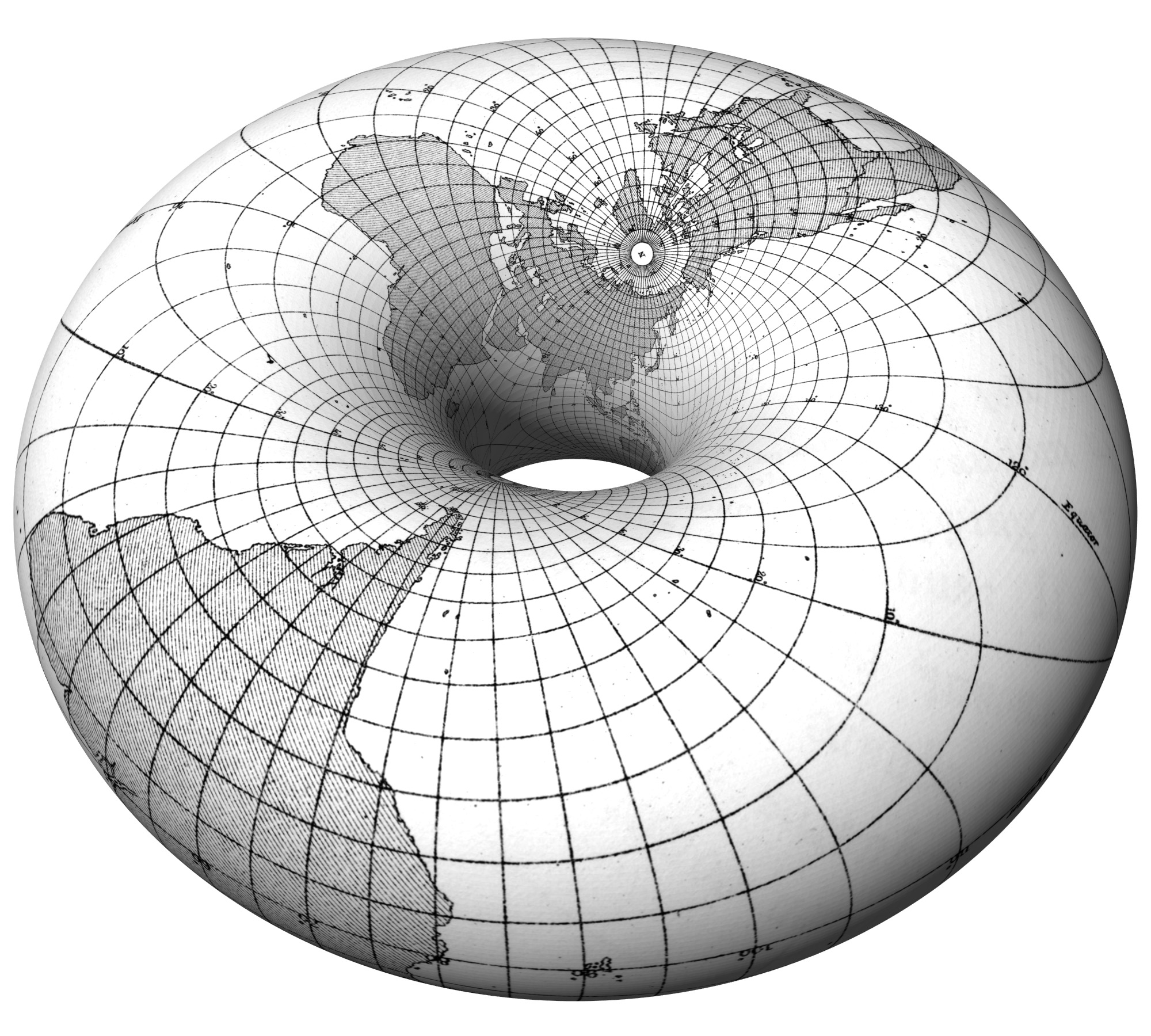}
\label{Fig:Peirce_torus}
} 
\caption{}
\label{Fig:Peirce}
\end{wrapfigure}

\mbox{}  %hack to make wrapfig not screw up
\vspace{-15pt}

\section*{Weierstrass and Schwarz-Christoffel} 
\label{Sec:Weierstrass}

%%% Question - is there a nice curve on the sphere which we can
%%% pullback via Weierstrass to the plane, rescale, push down, and
%%% then get the spiralling-type effect we see in Droste?  I think
%%% that the answer is sort of no...

The exponential and logarithm are just two of the many beautiful
flowers in the field of complex analysis.  More exotic ``elliptic''
functions can be used; as far as we are aware, the earliest
application to spherical images is due to Charles Sanders Peirce, in
1879~\cite{Peirce79} (see also \cite{Adams25}).
\reffig{Peirce_figure} shows Peirce's \emph{quincuncial projection};
in \reffig{Peirce_torus} we wrap it around the square
torus\footnoteB{\url{https://skfb.ly/NJRx}}.
%%% https://sketchfab.com/models/add7b316976c4bcb9ef83b0f7f055c7a

Consider the Weierstrass $\wp$--function; we
refer to~\cite[Chapter~7]{Ahlfors66} for an excellent and short
introduction.  As a series the function for the square lattice is:
\[
\wp_i(z) =
  \frac{1}{z^2} +
  \sideset{}{'}\sum \left( \frac{1}{(z - w)^2} - \frac{1}{w^2}\right).
\]
The sum ranges over the non-zero Gaussian integers $w \in
\ZZ[i]$. 
%%% We also rescale $\wp$ to arrange $\wp(1/2) = 1$.
It is a non-trivial exercise to check that $\wp_i(z + 1) = \wp_i(z +
i) = \wp_i(z)$.  That is, the Weierstrass $\wp$--function is
\emph{doubly periodic}.  In comparison, the exponential function is
only singly periodic: $\exp(z + 2\pi i) = \exp(z)$.  The above series
converges very slowly; for image processing we instead implement
$\wp_i$ using theta-functions~\cite[page~132]{McKeanMoll97}.

\begin{figure}[htbp]
\centering

\subfloat[The two-fold branched covering maps from the torus to the
  sphere by ``folding'' the torus around the red ``skewer''. The four
  skewered points of the torus become the four red dots on the
  sphere.]
{
\includegraphics[width=0.54\textwidth]{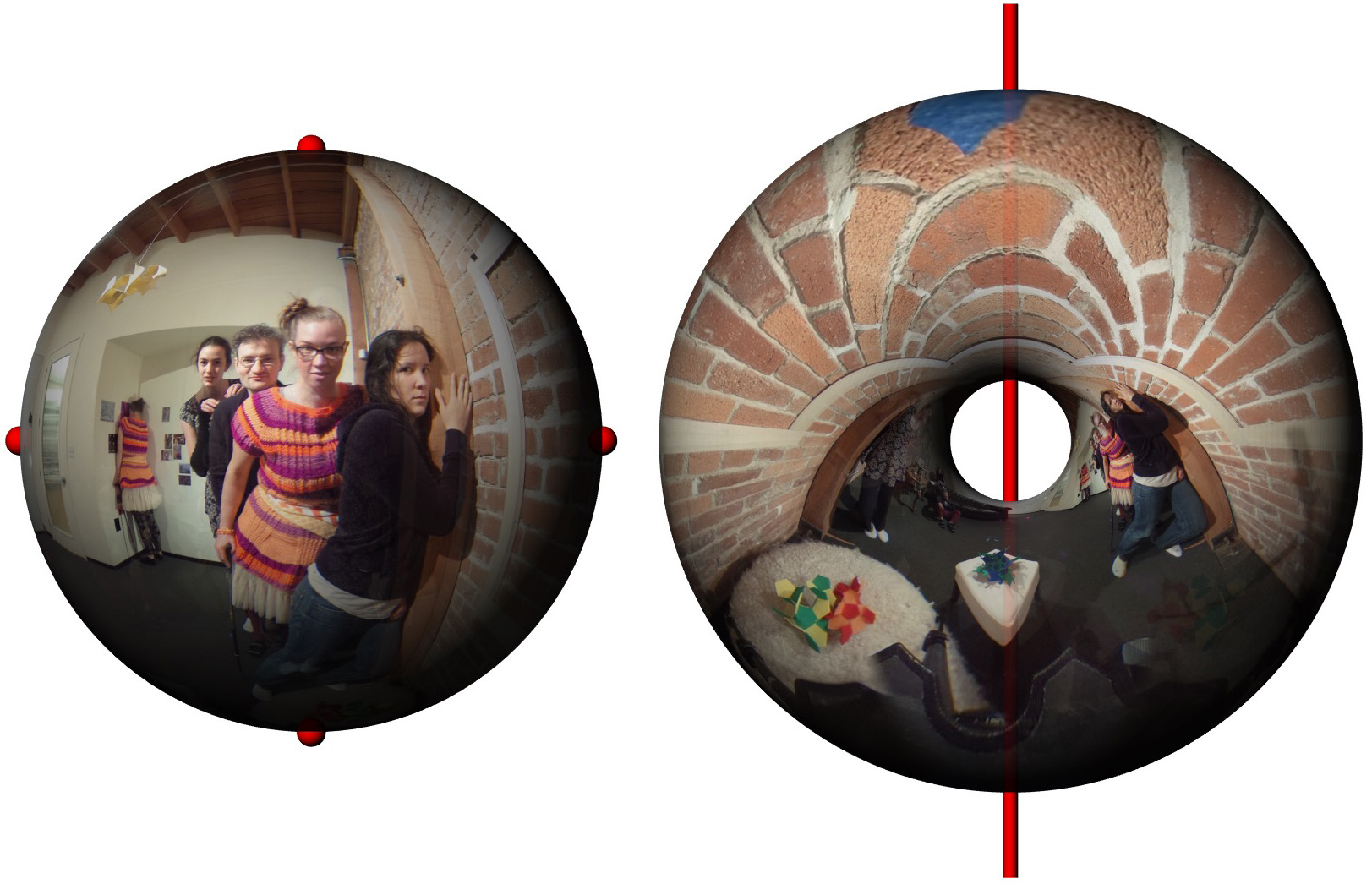}
\label{Fig:sphere_torus_skewer}
} 
\subfloat[Cutting open the torus yields a square that tiles the plane.]
{
\includegraphics[width=0.32\textwidth]{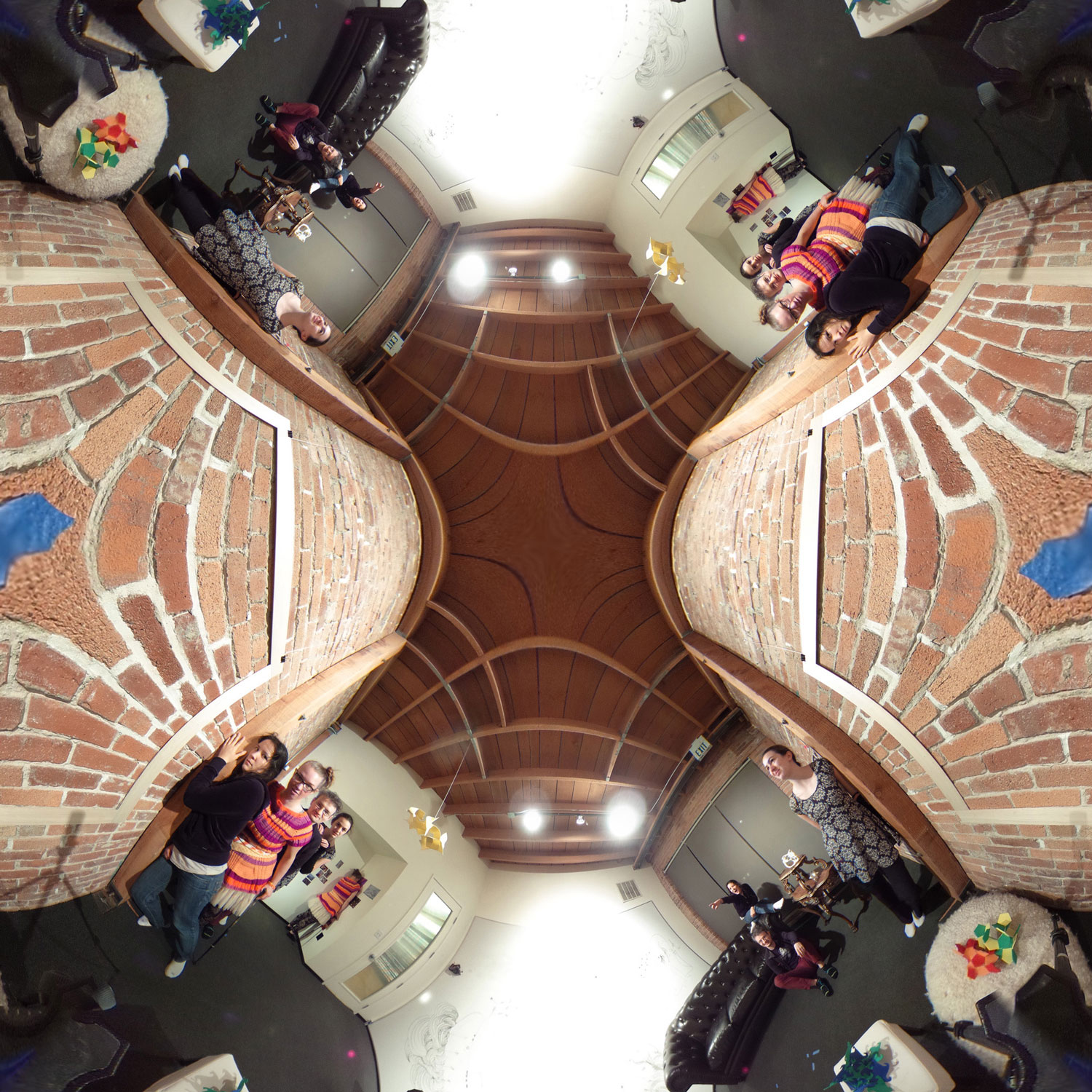}
\label{Fig:weierstrass_square1}
} 

% Note that: 
% 0 = ceiling
% inf = tripod
% 1 = back of room
% i = Vi
% -1 = blue tape
% -i = Emily
% Yay!

\subfloat[A fundamental domain after rescaling by $z \mapsto (1 + i)
  \cdot z$.]
{
\includegraphics[height=2in]{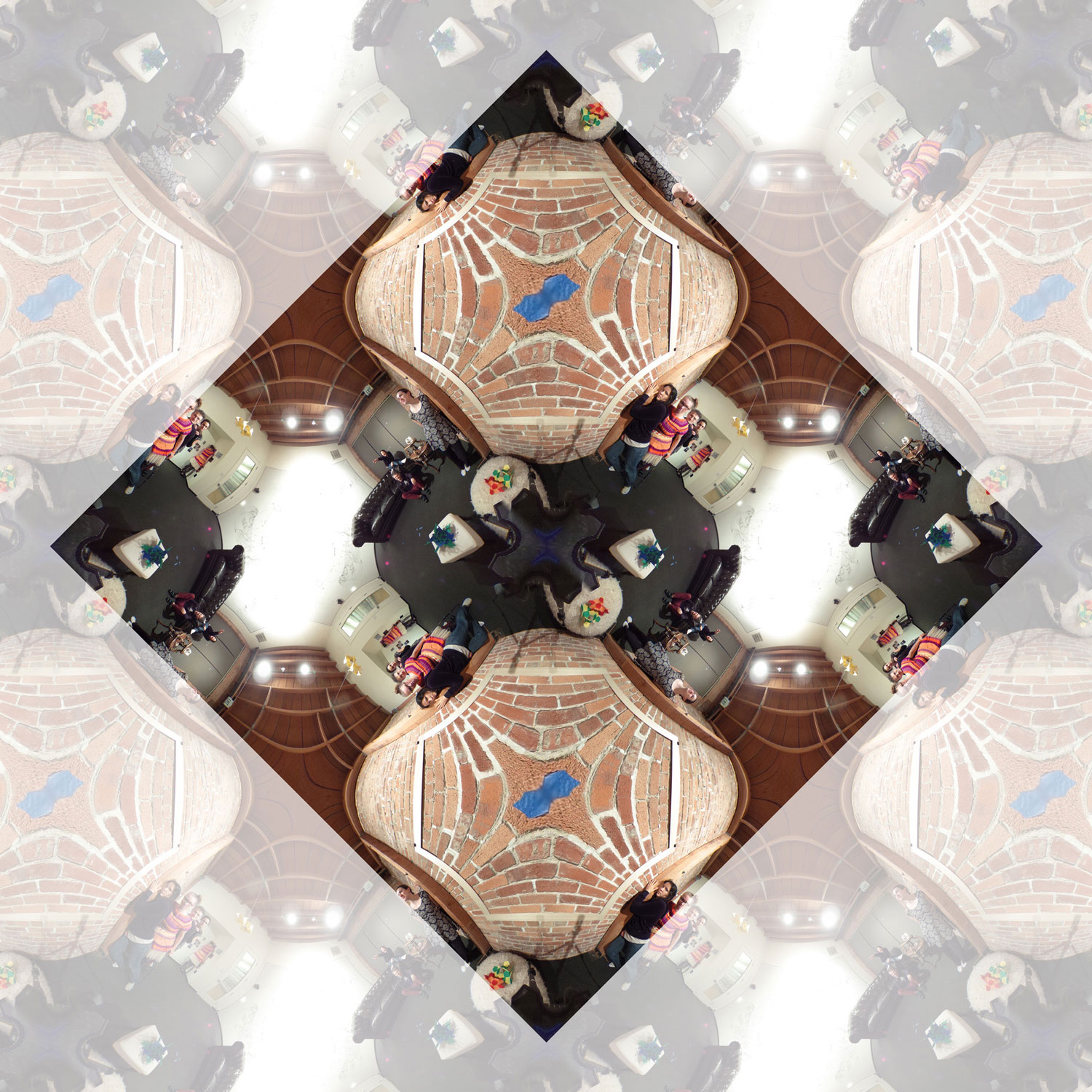}
\label{Fig:weierstrass_square2}
}
\subfloat[Map down to the sphere again via Schwarz-Christoffel.  The
  composition is the rational map $z \mapsto \frac{i}{2} (-z + 1/z)$.]
%%% Latt\`es - see Milnor p.71, problem 7f
{
\includegraphics[height=2in]{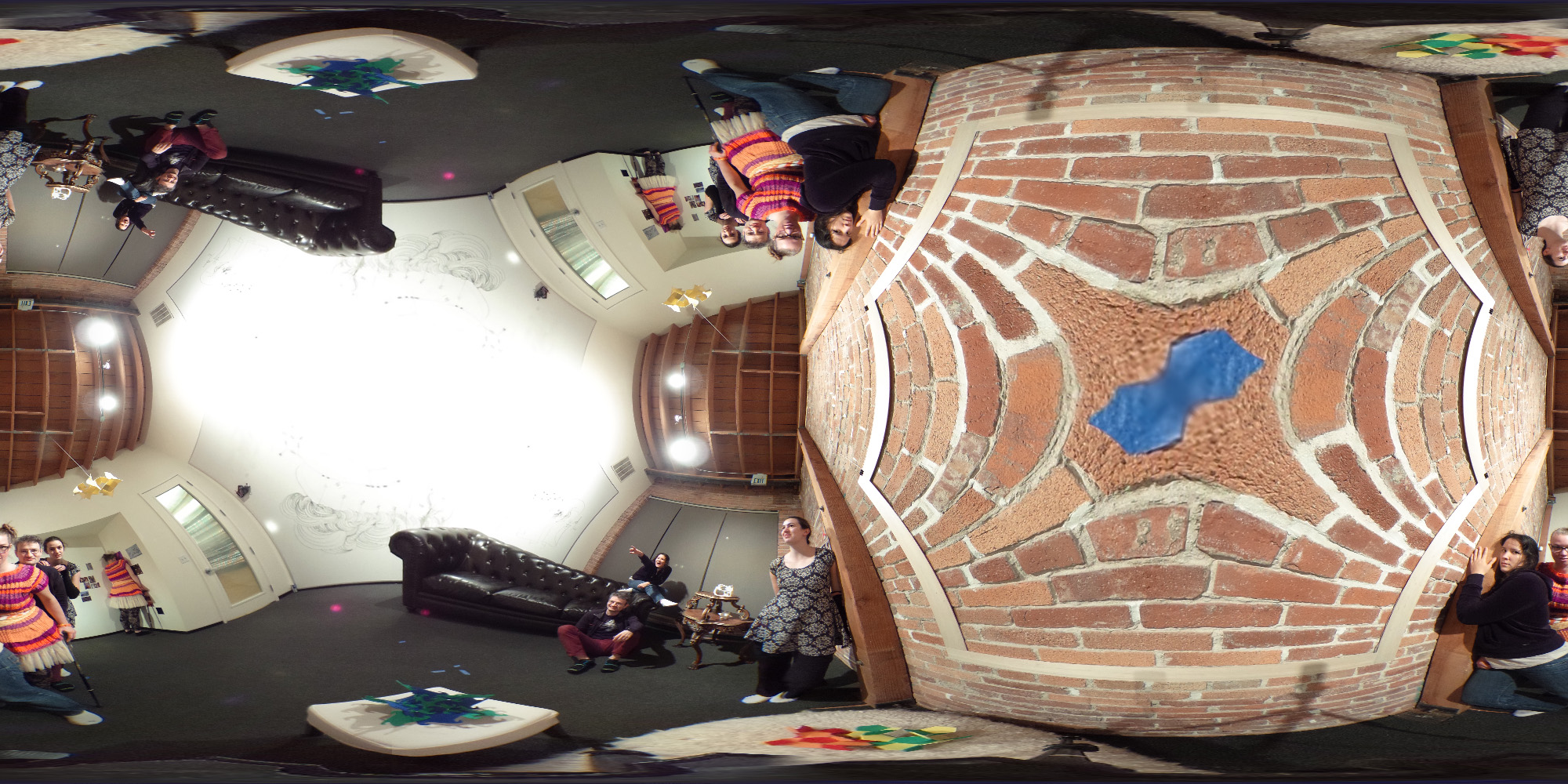}
\label{Fig:schwarz_christoffel_x(1+i)_weierstrass_2000}
}

\subfloat[The result if we instead use $z \mapsto 2 \cdot z$. The
  corresponding rational function is $z \mapsto
  \frac{(z^2+1)^2}{4z(z^2-1)}$.]
%%% Also Latt\`es
{
\includegraphics[width=0.48\textwidth]{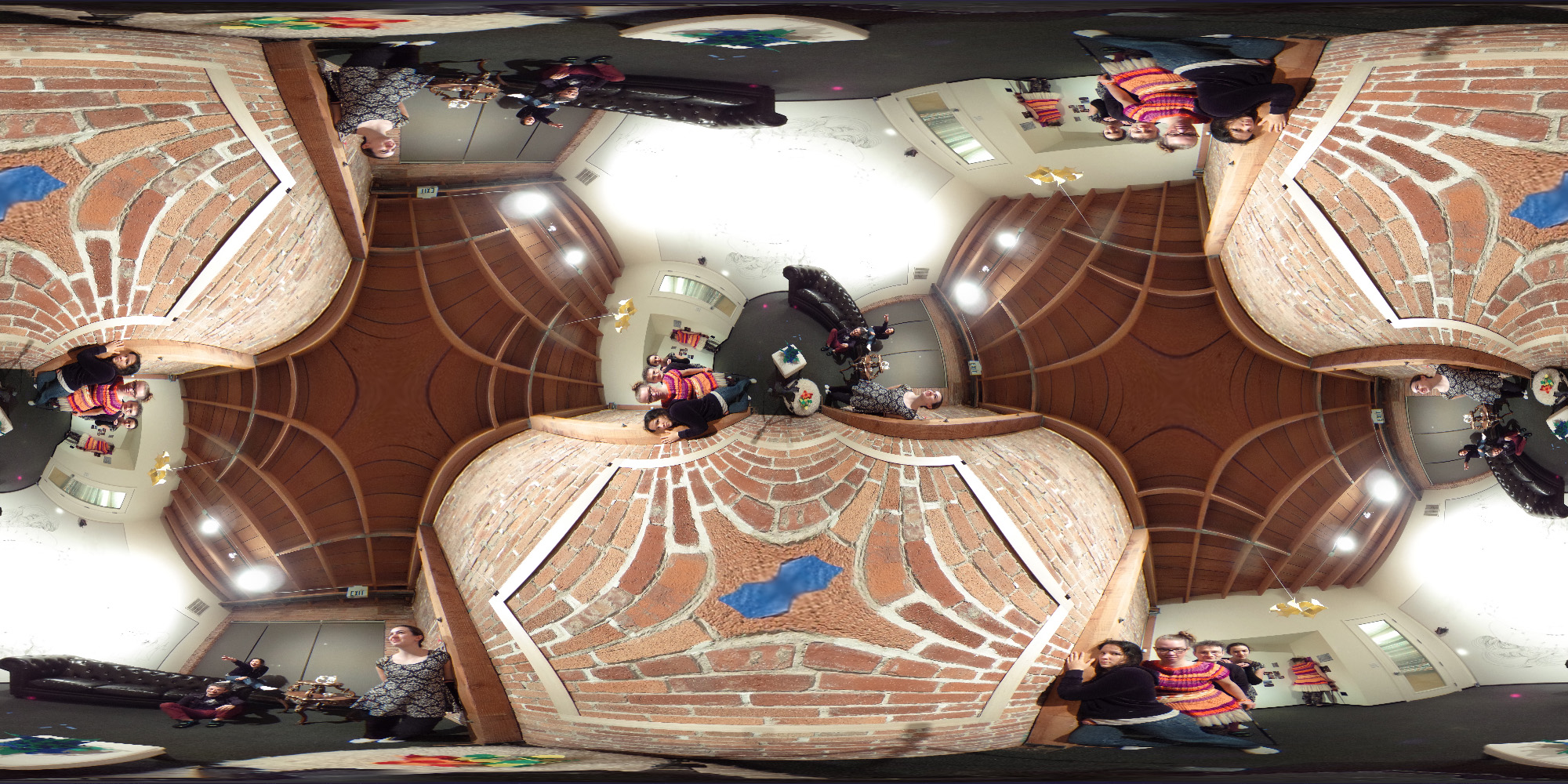}
\label{Fig:schwarz_christoffel_x2_weierstrass_2000}
}
\subfloat[The result if we instead use $z \mapsto (2+i) \cdot z$.  Now
  we obtain the rational map $z \mapsto z \frac{((-1 + 2i) +
    z^2)^2}{(-i + (2 + i)z^2)^2}$.]
%%% Also Latt\`es 
{
\includegraphics[width=0.48\textwidth]{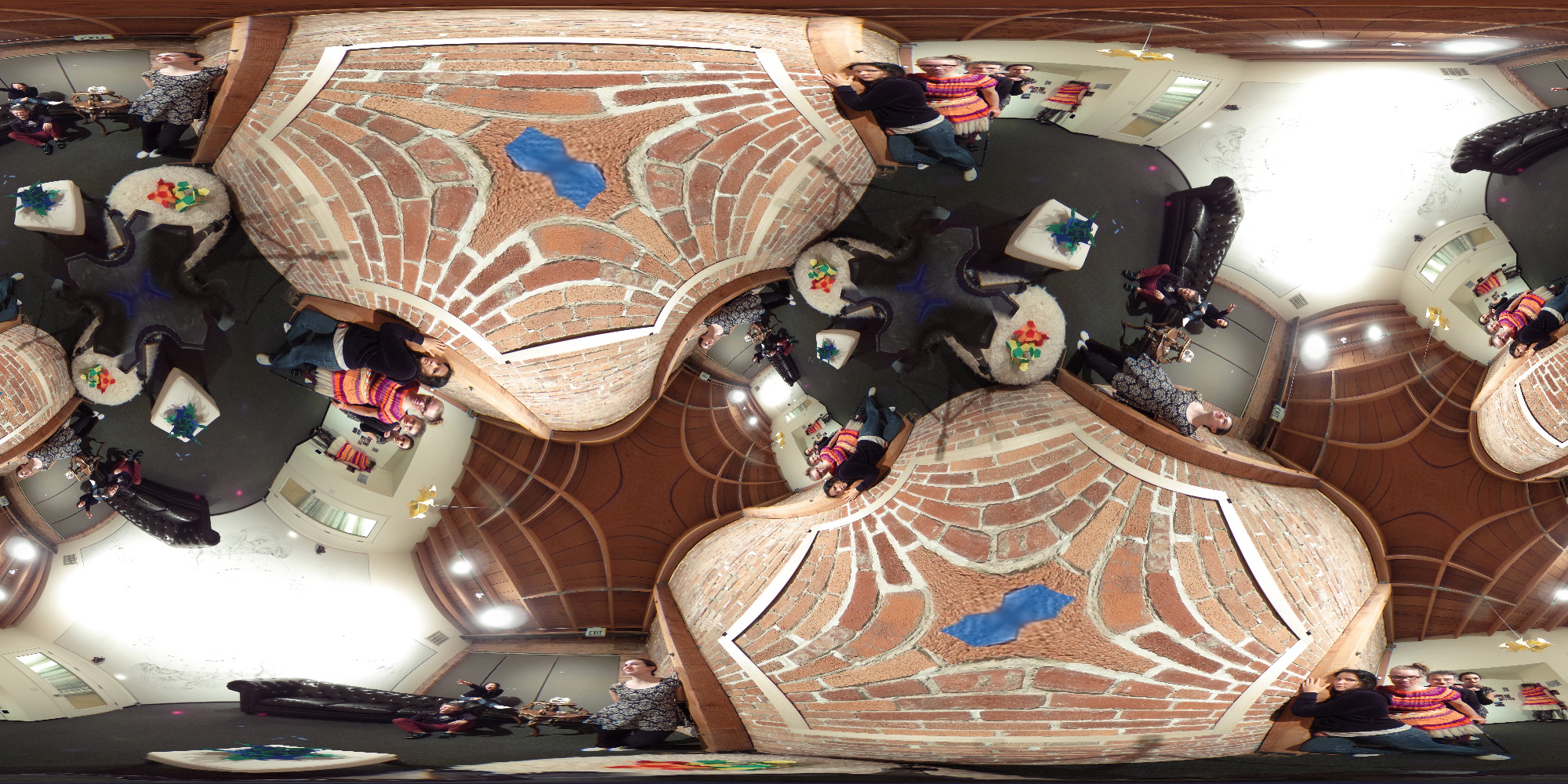}
\label{schwarz_christoffel_x(2+i)_weierstrass_2000}
} 

\caption{Images produced using Weierstrass and Schwarz-Christoffel maps.}
\label{Fig:weierstrass_maps}
\end{figure}

Since $\wp_i$ is doubly periodic, we think of it as first mapping the
plane $\CC$ to the square torus $\TT$, which then maps to the Riemann
sphere $\RS$ via a branched double-cover.  So, we start with our
standard spherical image (\reffig{sphere_torus_skewer},
left\footnoteB{\url{https://skfb.ly/NKox}}).  We pullback to $\TT$ and
obtain a toroidal image (\reffig{sphere_torus_skewer},
right\footnoteB{\url{https://skfb.ly/NKpq}}).  Note that the toroidal
image contains two copies of the original, and has four branch points.
%%% Compare to the version using z^2.
Cutting $\TT$ open we obtain \reffig{weierstrass_square1};
% as promised by the title of our paper this is
a square containing two copies of the original, spherical, image.
This is the unit cell of a tiling of $\CC$, obtained by pulling back
via $\wp_i$.  
%%% The square torus is cut into four squares and each is sent to a
%%% hemisphere of $\RS$.  This uses the Schwarz reflection principle.

Just as the exponential function has its logarithm, the Weierstrass
function $\wp_i$ has an inverse. This map, denoted $\SC_4$, takes the
disk to the square.
%%% Well, when you restrict it to the disk, it does. 
This, then, is a \emph{Schwarz-Christoffel}
function~\cite[Section~6.2.2]{Ahlfors66}.  In general, these functions
are given by difficult integrals, but for regular $n$--gons there is a
very pretty expression in terms of the hypergeometric
function~\cite[Exercise~5.19]{McDougallEtAl12}: % page 332
\[
\SC_n(z)
= \int_0^z \frac{dw}{(1 - w^n)^{\frac{2}{n}}}
= z \cdot {}_2F_1\left(\frac{1}{n}, \frac{2}{n}; 1 + \frac{1}{n}; z^n\right).
\]

\addtocounter{footnote}{4}  % get footnote and footnoteB on the same number

We are now ready to ``twist'', in similar spirit to the twisted Droste
effect.  We pullback the tiling in $\CC$ via the map $z \mapsto (1 +
i) \cdot z$.  A unit cell for this finer tiling is shown in
\reffig{weierstrass_square2}.  We pull this back to $\RS$ using the
map $\SC_4$ and obtain
\reffig{schwarz_christoffel_x(1+i)_weierstrass_2000}.  Pulling back
(in $\CC$) by other Gaussian integers gives other interesting effects;
see Figures~\ref{Fig:schwarz_christoffel_x2_weierstrass_2000}, and
\ref{schwarz_christoffel_x(2+i)_weierstrass_2000}.

\begin{figure}[htbp]
\centering

\subfloat[Glue opposite sides to obtain the hexagonal torus.]
{
\includegraphics[width=0.3\textwidth]{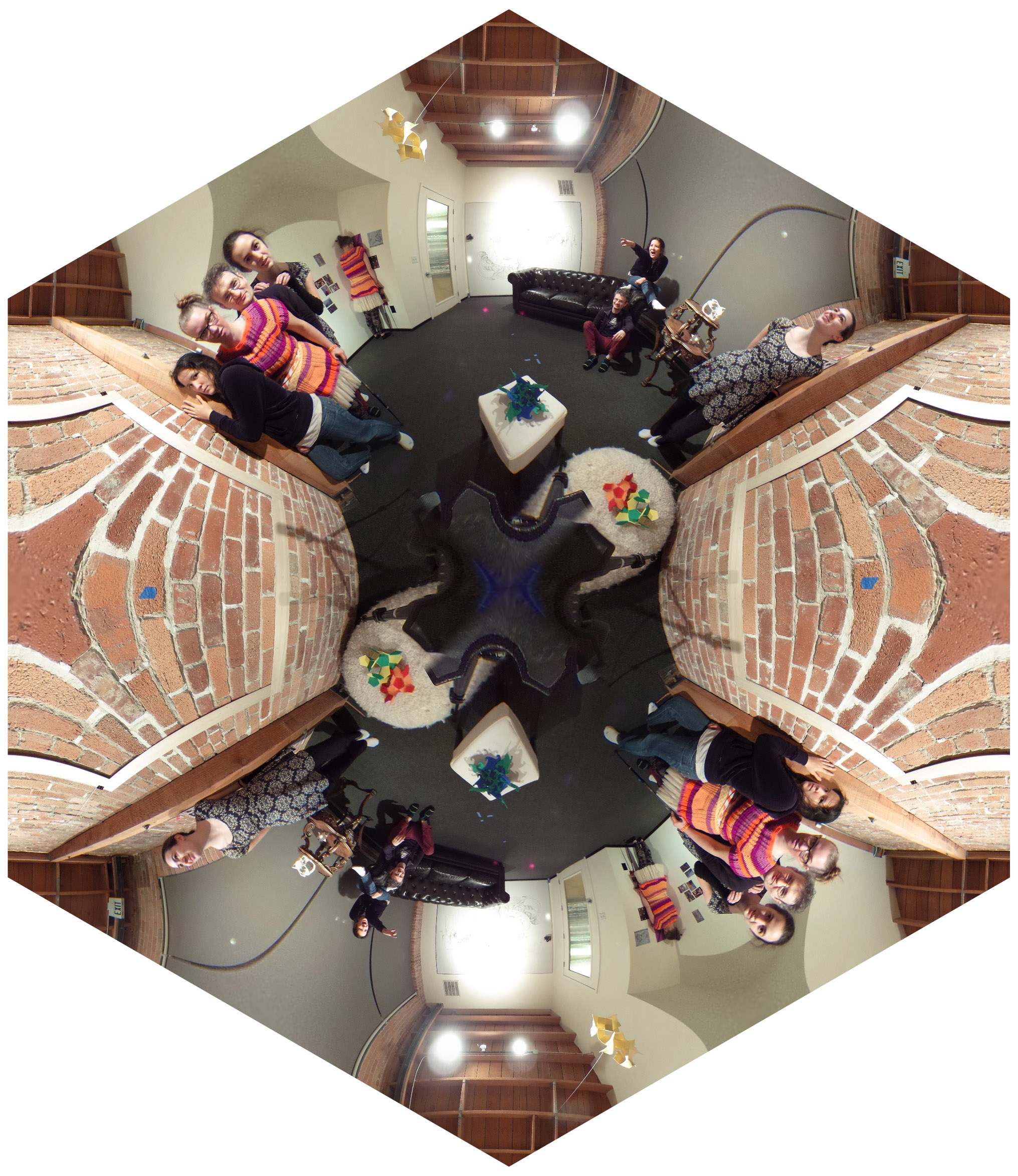}
\label{Fig:weierstrass_hex}
} 
\quad
\subfloat[A fundamental domain after rescaling by $z \mapsto (1 + \omega) \cdot z$.]
{
\includegraphics[width=0.4\textwidth]{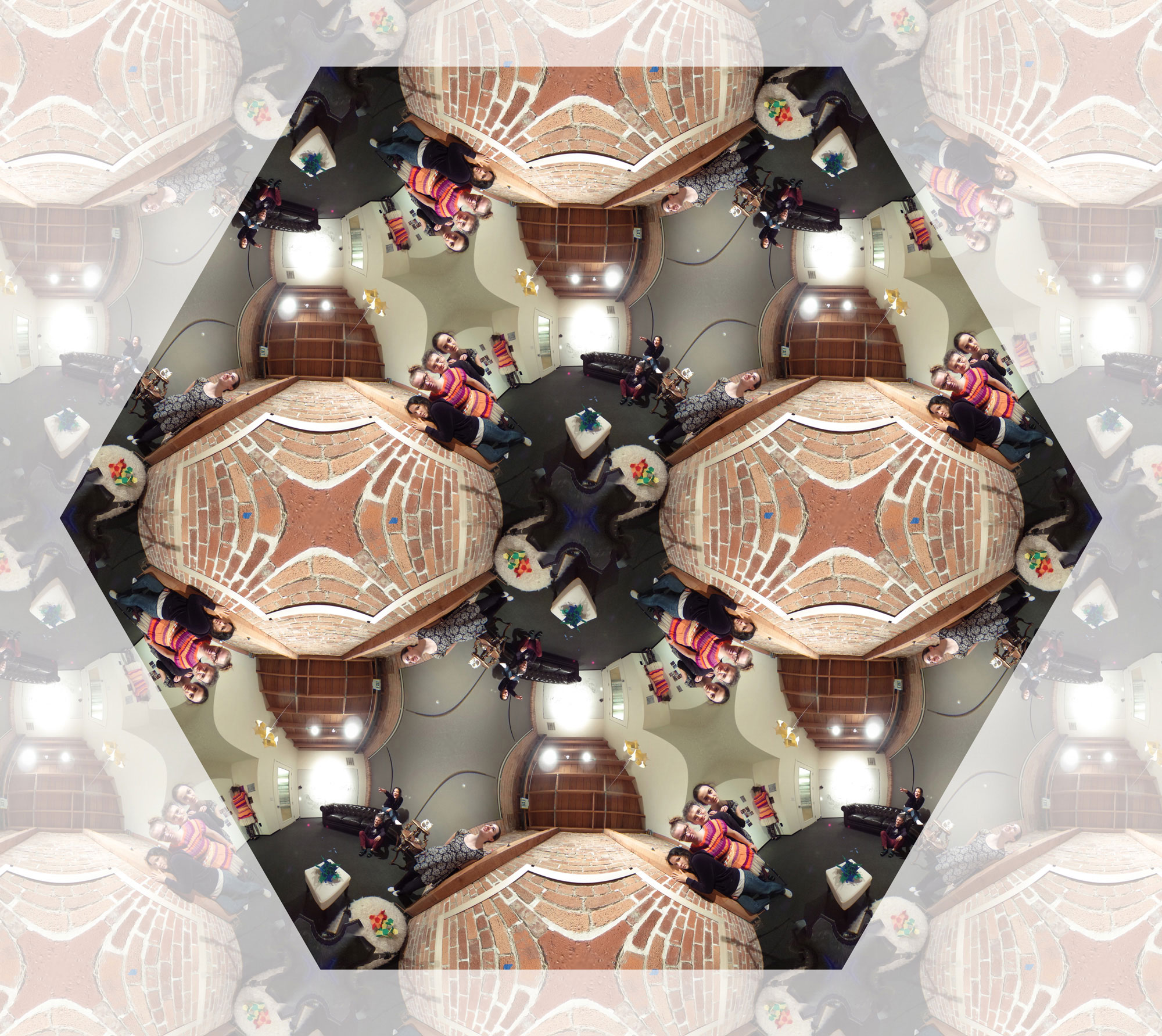}
\label{Fig:weierstrass_hex_1+omega}
} 

\subfloat[The result of pulling up and pushing down is the rational
  function is $z \mapsto \frac{z^3+\sqrt{2}}{3 \omega \cdot z^2}$.]
{
\includegraphics[width=0.7\textwidth]{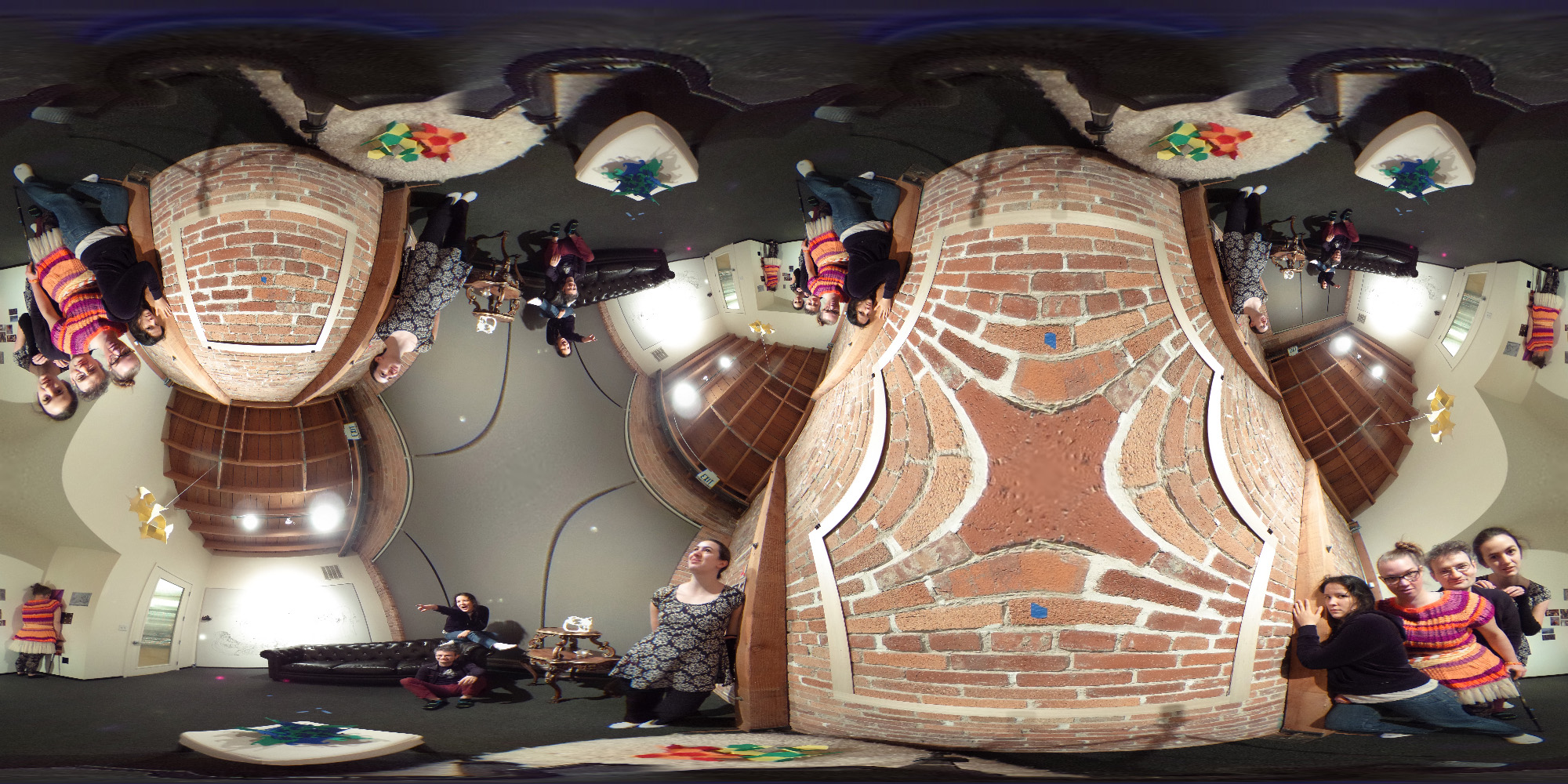}
\label{Fig:triple_cover}
}

\caption{Images produced with the hexagonal torus replacing the square
  torus.}
\label{Fig:weierstrass_maps_hex}
\end{figure}

It is well-known that gluing opposite sides of a square produces a
torus.  Less familiar is the fact that a torus also results from
gluing opposite sides of a hexagon.  We can now perform similar
transformations to those above using the hexagonal torus.  We pull our
standard image \reffig{elevr_image} back using the Weierstrass
function $\wp_\omega$ where $\omega$ is the usual sixth root of unity
-- that is, the sum is over the lattice $\ZZ[\omega]$.  We now
pullback by $z \mapsto (1 + \omega)z$ and then by the appropriate
Schwarz-Christoffel function.  See \reffig{weierstrass_maps_hex}.

%%% Rational functions:
%%% z \mapsto \frac{i}{2} (-z + 1/z)
%%% 
%%% z \mapsto \frac{z^4+2z^2+1}{4z^3-4z} aka z \mapsto
%%% \frac{(z^2+1)^2}{4z(z^2-1)}
%%%
%%% z \mapsto z\frac{-z^4 + (2-4i)z^2 + (3+4i)}{(-3-4i)z^4 +
%%% (-2+4i)z^2 + 1} aka z \mapsto z \frac{((-1 + 2i) + z^2)^2}{(-i +
%%% (2 + i)z^2)^2}

In all cases the overall map from $\RS$ to $\RS$ is conformal, apart
from a finite set of points.  Thus it is in fact a \emph{rational
  map}~\cite[Section~4.3.2]{Ahlfors66}.  It is far faster to use this
rational function rather than the composition of elliptic and
hypergeometric functions.  It is possible to find the rational
function synthetically~\cite[Chapter~7]{Milnor06}, but we found it
simpler to proceed numerically, as follows.  We must find the
coefficients $\{a_i, b_i\}$ of a rational function $f(z) =
(\sum_{i=0}^n a_iz^i)/(\sum_{i=0}^n b_iz^i)$ of known degree $n$.
Since $f$ is given above as a composition we can sample $f$ at a
number of points $z_j$, getting the results $w_j = f(z_j)$.  Each
sample gives a row
\[
(z_j^n, z_j^{n-1}, \ldots, z_j, 1, -w_j z_j^n, -w_j z_j^{n-1}, \ldots,
-w_j z_j, -w_j)
\]
of a matrix $M$ with the property that $M \cdot (a_n, a_{n-1}, \ldots,
a_1,a_0, b_n, b_{n-1}, \ldots, b_1,b_0)^T = 0$.  With enough samples,
$M$ has a one-dimensional kernel, which can be found using the 
singular value decomposition method.

\begin{figure}[t]
\begin{center}
\subfloat[The input image.]
{
\includegraphics[width=0.48\textwidth]{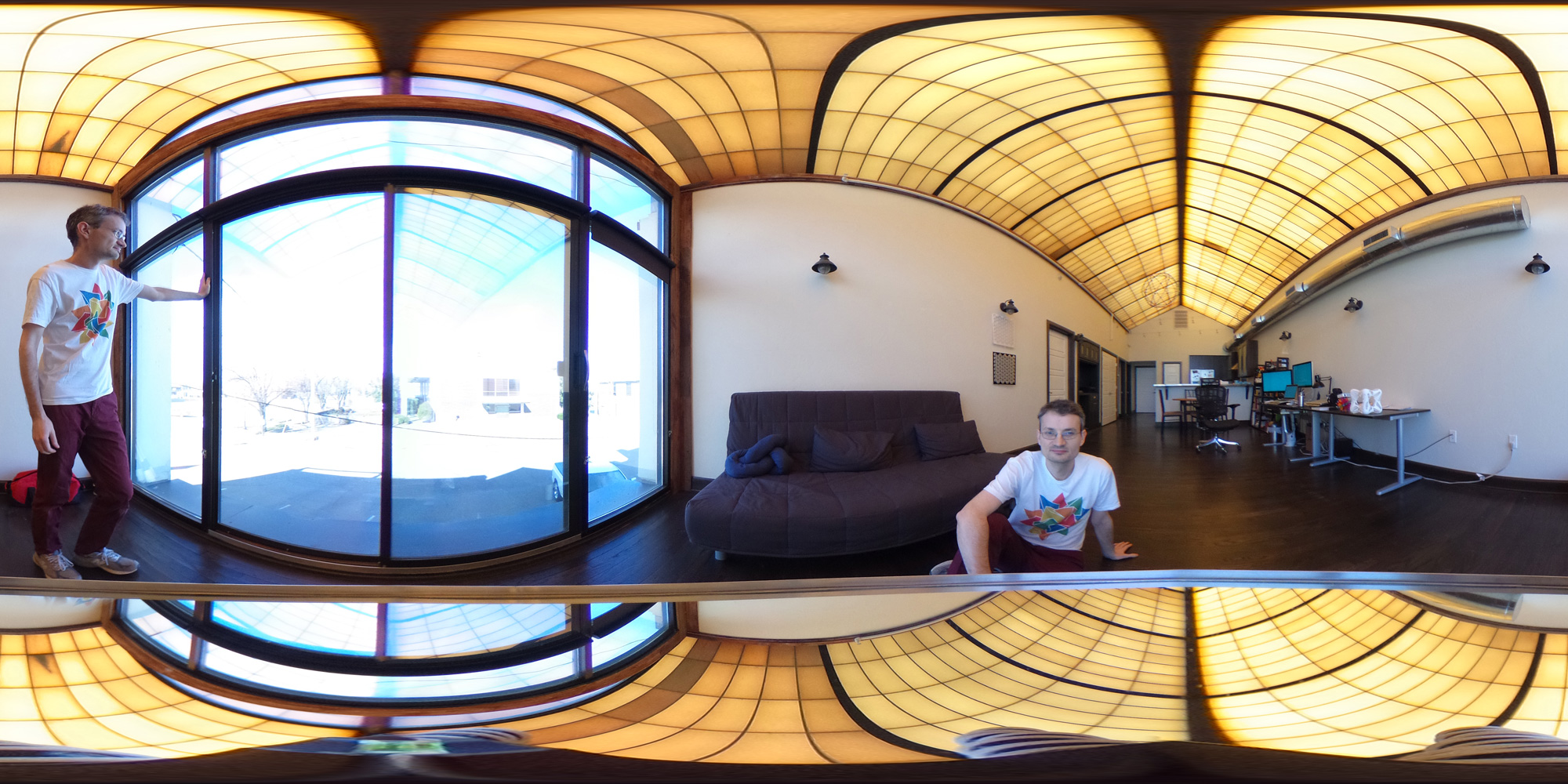}
\label{Fig:schottky_input}
}
\subfloat[The disks $D_a$, $D_A$, $D_b$, $D_B$ in the Riemann sphere.]
{
\labellist
\small\hair 2pt
\pinlabel {$D_a$} at 1348 1385
\pinlabel {$D_A$} at 3083 1479
\pinlabel {$D_b$} at 2690 319
\pinlabel {$D_B$} at 4436 2227
\endlabellist
\includegraphics[width=0.48\textwidth]{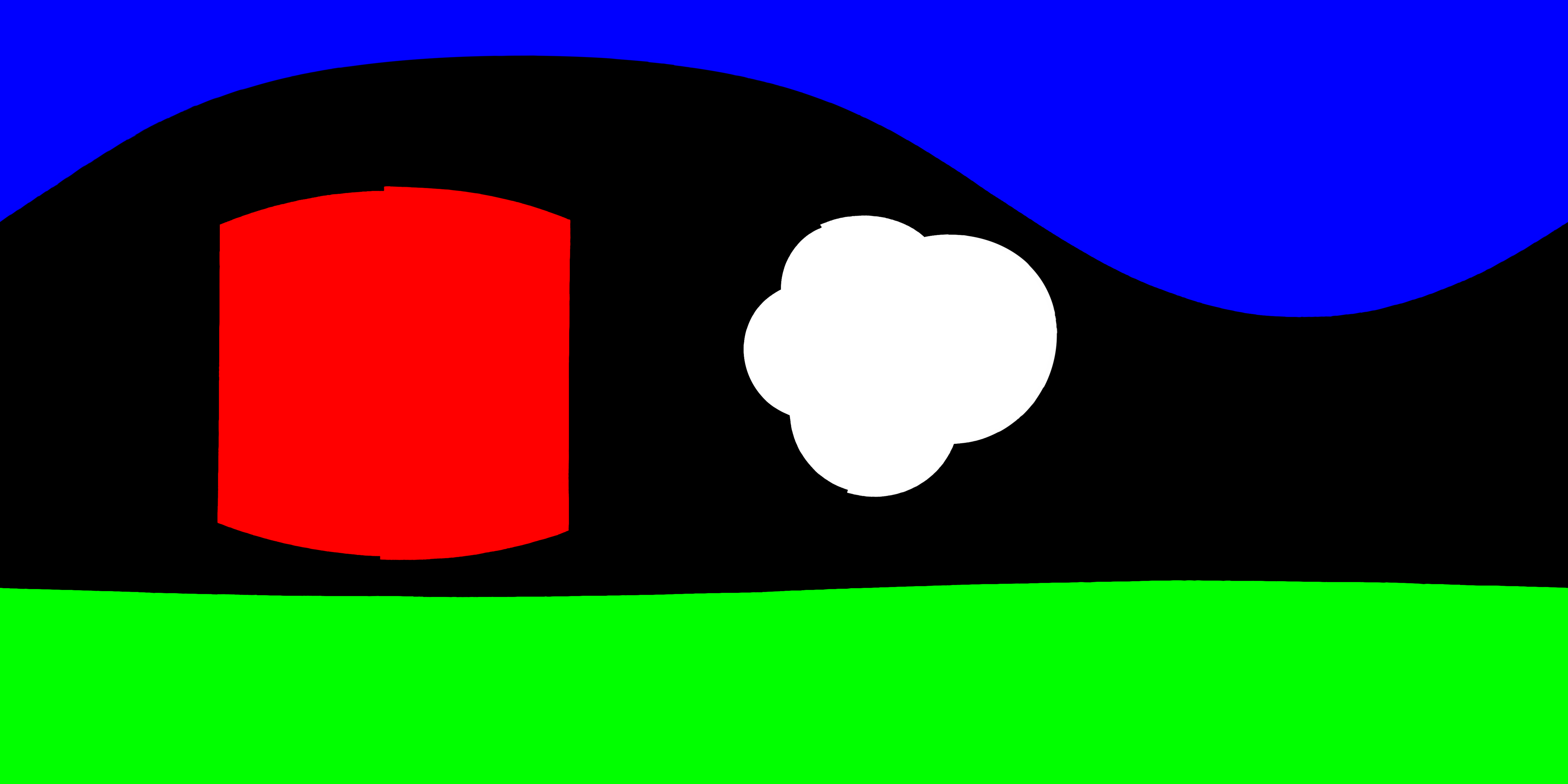}
\label{Fig:schottky_mask}
}
\end{center}

\subfloat[A spherical Schottky image.]
{
\includegraphics[width=\textwidth]{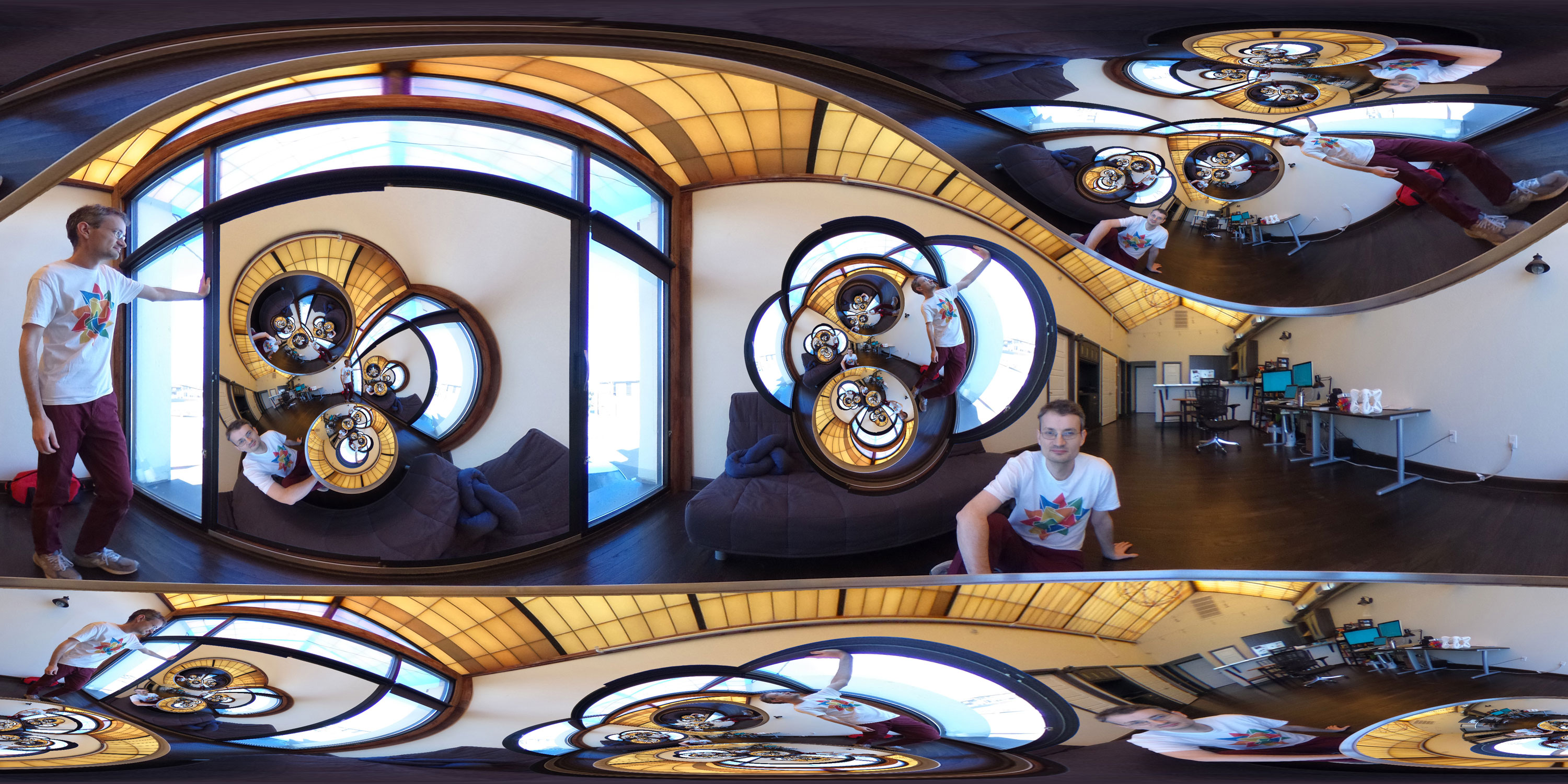}
\label{Fig:schottky_result}
}
\caption{A spherical double Droste effect, using a Schottky group.}
\label{Fig:schottky_droste}
\end{figure}

\section*{Schottky groups}
\label{Sec:Schottky}

Suppose that $a$ and $b$ are hyperbolic (that is, zooming) M\"obius
transformations.  Suppose that $D_a$, $D_A$, $D_b$, $D_B$ are four
closed disjoint disks in $\RS$
% so that $a(D_A) \cap D_a = \bdy D_a$ and $b(D_B) \cap D_b = \bdy D_b$
so that $a$ maps the interior of $D_A$ onto the exterior of $D_a$, and
similarly for $b$.  Then the group generated by $a$ and $b$ is called
a two-generator \emph{Schottky group}~\cite[page~98]{MumfordEtAl02}.
%%% The general definition has indices - it is a bit too heavy for this
%%% style of paper.  Note that Schottky implies free, by the ping-pong
%%% lemma.
%%%
%%% Classical Schottky - the disks must be round. (ie ``isometric
%%% circles'').  If they are tangent then fun stuff can happen - like
%%% the limit set can become a closed curve.  If the circles intersect
%%% a lot, then you can start to get indiscrete groups.  If they
%%% intersect even more, then you sometimes get unfaithful groups
%%% which are discrete - that is, a relation shows up.  This is how
%%% Jorgenson discovered the infinite cyclic cover of the figure eight
%%% knot group.

Schottky groups can be used to generate impressive images; for a
richly illustrated introduction to the underlying mathematics
see~\cite{MumfordEtAl02}. David Gu has also experimented with applying
Schottky reflection groups to photographs\footnote{See
  \url{http://www3.cs.stonybrook.edu/~gu/lectures/lecture_1_Escher_Droste_Effect.pdf}.}.
We now discuss how to apply these ideas to spherical images.

We begin with an input image \reffig{schottky_input}.  We must choose
the positions of the disks $D_a$, $D_A$, $D_b$, and $D_A$.  In the
final image these will contain zoomed copies of (part of) the input
image.  For $D_a$ we choose the window; for $D_b$ we choose the large
round mirror lying below the camera, on which the tripod is standing.
We trace over $D_a$ and $D_b$ in Photoshop to make a mask image in
which the window is red and the mirror green, as shown in
\reffig{schottky_mask}.  We now choose two hyperbolic M\"obius
transformations, $A$ and $B$, and set $D_A = A(\RS - D_a)$, in white,
and $D_B = B(\RS - D_b)$, in blue.  We choose $A$ and $B$ so that the
all of the disks are disjoint, and no disk covers an important part of
the input image.  Let $a = A^{-1}$ and define $b$ similarly.

To generate the image\footnote{Also see an animated version:
  \url{https://www.youtube.com/watch?v=vtWtmTzGxd4}.} shown in
\reffig{schottky_result}, for each pixel $p$, we perform the following
routine.
\begin{enumerate}
\item
  Set $q = p$.
\item
  If $q$ lies in the black region of the mask, color $p$ the same
  as the color of $q$ and stop the routine.
\item
  Otherwise, if $q$ lies in $D_X$ then replace $q$ by $x(q)$ and go to
  step $2$.
\end{enumerate}

In general, a Schottky group can have more than two generators,
or indeed fewer.  Using just one generator recovers the straight
Droste effect; $\RS - (D_a\cup D_A)$ is the Droste annulus.
It is interesting to ponder how one might apply the twisted Droste
effect throughout a Schottky image, but that is a task for another
day.

\bibliographystyle{plain} 
\bibliography{spherical}

\end{document}